\documentclass[english,preprint]{aastex} \usepackage{natbib}
\usepackage{graphicx} 

\makeatletter

\def\lsim{\raisebox{-.4ex}{$\stackrel{<}{\scriptstyle \sim}$\,}}

\usepackage{babel}
\makeatother

\begin{document}

\title{Full Numerical Simulations of Catastrophic Small Body Collisions\altaffilmark{1}}
\shorttitle{Catastrophic Disruption of Small Bodies}
\shortauthors{Leinhardt \& Stewart}

\author{Zo\"e M. Leinhardt\altaffilmark{2,3,4}}

\affil{Department of Applied Mathematics and Theoretical Physics, University of Cambridge, Cambridge, CB3 0WA, U.K., z.m.leinhardt@damtp.cam.ac.uk}

\author{Sarah T. Stewart} \affil{Earth and Planetary Sciences
  Department, Harvard University, 20 Oxford St., Cambridge, MA 02138, sstewart@eps.harvard.edu}

\altaffiltext{1}{Accepted for publication in Icarus doi \#10.1016/j.icarus.2008.09.013}
\altaffiltext{2}{STFC Postdoctoral Fellow}
\altaffiltext{3}{Harvard College Observatory, Cambridge, MA, 02138}
\altaffiltext{4}{Corresponding author.}

\begin{abstract}
  The outcome of collisions between small icy bodies, such as Kuiper
  belt objects, is poorly understood and yet a critical component of
  the evolution of the trans-Neptunian region. The expected physical
  properties of outer solar system materials (high porosity, mixed
  ice-rock composition, and low material strength) pose significant
  computational challenges.  We present results from catastrophic
  small body collisions using a new hybrid hydrocode to $N$-body code
  computational technique. This method allows detailed modeling of
  shock propagation and material modification as well as gravitational
  reaccumulation.  Here, we consider a wide range of material
  strengths to span the possible range of Kuiper belt objects.  We
  find that the shear strength of the target is important in
  determining the collision outcome for 2 to 50-km radius bodies,
  which are traditionally thought to be in a pure gravity regime. The
  catastrophic disruption and dispersal criteria, $Q_D^*$, can vary by
  up to a factor of three between strong crystalline and weak
  aggregate materials.  The material within the largest reaccumulated
  remnants experiences a wide range of shock pressures.  The dispersal
  and reaccumulation process results in the material on the surfaces
  of the largest remnants having experienced a wider range of shock
  pressures compared to material in the interior.  Hence, depending on
  the initial structure and composition, the surface materials on
  large, reaccumulated bodies in the outer solar system may exhibit
  complex spectral and albedo variations. Finally, we present revised
  catastrophic disruption criteria for a range of impact velocities
  and material strengths for outer solar system bodies.
\end{abstract}


\section{Introduction}

Kuiper belt objects (KBOs) are some of the least altered bodies in the
solar system. Hence, the properties of KBOs and their cometary
fragments may faithfully reflect the composition and physical
structure of the volatile-rich planetesimals that accreted into the
larger planets, thus providing insights into the conditions in the
solar nebula. Astronomical observations of KBOs have revealed
variations in the surface composition of the largest bodies
\citep{Brown08}. In addition, the colors of KBOs may suggest division
into subgroups with different dynamical histories
\citep{Doressoundirum01,Jewitt01,Doressoundirum05}. Bulk densities and
inferences about internal structure may also be derived from remote
sensing \citep{Noll08,Stansberry08}. One of the few modification
processes that have acted upon KBOs is mutual collisions, which may
alter both composition and internal structure \citep{Leinhardt08}. In
this work, we develop methods to address the extent to which KBOs have
been modified by mutual collisions since the last stages of planetary
accretion.

Over the age of the solar system, the surfaces of KBOs have been
modified by high energy particles, micrometeorite bombardment, impact
cratering events and mutual collisions, and possible endogenic
resurfacing \citep{Stern03}. While space weathering processes tend to
darken and redden the surfaces of asteroids \citep{Chapman04}, the net
effect in the Kuiper belt has been difficult to quantify, partly
because of the unknown surface composition and chemistry. However, the
observed high albedos and range of colors demonstrate that space
weathering does not dominate the surface properties of KBOs.
Similarly, the role of endogenic processes is poorly constrained.
Unlike the largest bodies in the asteroid belt, the effects of
radioactive decay are minimal or negligible in the thermal evolution
of Kuiper belt objects because of the longer accretion time scales. As
a result, few classical KBOs may have differentiated or ever supported
cryovolcanic processes \citep{Merk06}.

The cumulative effects of impact cratering and catastrophic collisions
in the Kuiper belt have also been difficult to quantify. KBOs
currently reside in a low number density region of the solar system.
So, although the eccentricities and inclinations of the bodies are
relatively high in comparison to the major planets, the present day
mutual collision probabilities are low \citep{Durda00}. However, there
is abundant evidence for a significant collisional history in the
Kuiper belt from observations of dust production, the size
distribution of bodies, and the low present-day total mass of KBOs
\citep{Leinhardt08}. Based on models of the dynamical history of the
Kuiper belt, most KBOs with diameters $< 100$ km have suffered large
collisions over their lifetime \citep{Durda00,Farinella96}. It is
these large collisions with other KBOs that may have significantly
changed the internal structure and/or contributed to devolatilization
of the primordial bodies.  In order to assess the effects of
collisions on the properties of KBOs, the outcome of collisions must
be related to observable parameters, e.g., surface composition,
colors, and densities. The range of possible collision outcomes
depends on the initial properties of KBOs (mass, density, composition,
and internal structure) and the impact conditions (velocity, impact
angle). Single impact events are generally expected to brighten the
surface by excavating through the space-weathered surface into fresh
volatile-rich layers. However, if KBOs are highly porous and mixtures
of ices and refractory materials, small cratering events may result in
compaction of surface layers and possibly net devolatilization, which
could darken the surface \citep{Leinhardt08,Housen03}. The cumulative
effects of individual impact cratering events will depend on the
dynamical history in the Kuiper belt. More energetic impact events
lead to total disruption of the bodies and gravitational
reaccumulation into many smaller fragments. The fragments may or may
not reflect the general physical properties of the original target.

The outcome of collision events fall into three general categories:
cratering, shattering, and dispersal. Cratering events affect only the
surface of the target. Shattering events break a coherent target into
fragments. Dispersal events not only shatter a coherent target, but
also cause fragments to escape the largest remnant. Each type of
collision outcome may produce changes in the physical properties of
small bodies in the outer solar system \citep{Leinhardt08}, but the
most energetic dispersal events produce potentially observable changes
in the internal structure and surfaces of the reaccumulated bodies.

Shattering and dispersal impacts are often discussed with respect to
the catastrophic shattering criteria and the catastrophic disruption
and dispersal criteria, $Q_S^*$ and $Q_D^*$, respectively. The $Q_S^*$
curve describes the energy per unit mass of the target delivered by
the projectile such that the largest intact fragment is half the mass
of the original target. In comparison, the $Q_D^*$ curve describes the
specific energy necessary from an impact such that the largest remnant
is half the mass -- either as a single intact fragment or as a
reaccumulated rubble pile. In the strength regime (target radii,
$R_T$, $\lsim 1$ km), $Q_S^*$ and $Q_D^*$ are equivalent, and $Q_S^*$
decreases with increasing target size due to the effects of strain
rate and length scale on the tensile strength \citep{Grady80,
  Housen90, Housenb99}.  The slope of the $Q_S^*$ curve changes sign
when the lithostatic pressure is comparable to the tensile srength of
the target ($R_T > 10$'s to 100~km) \citep[e.g.,][]{Melosh97}.
However, the slope of the $Q_D^*$ curve changes sign at a smaller
target radii ($\sim 1$~km), when gravitational reaccumulation becomes
important in determining the mass of the largest post-collision
remnant defining the transition from the strength to the gravity
regime. Here, we focus on the gravity regime.  The intercept between
the two regimes defines the weakest material in the system, which is
important for the collisional evolution of a population.

To accurately simulate collisions between Kuiper belt objects, an
ideal numerical technique must include the capability (i) to use
multiple equations of state in compositional mixtures, (ii) to model
compaction of pore spaces, and (iii) to model a variety of internal
structures.  In this work, we develop a new technique that can more
accurately model the possible range of material properties of KBOs
than techniques used in the past. In this paper, we focus on the
effects of material strength on $Q_D^*$, leaving the investigation of
porosity and mixed materials for future work.

\section{Numerical Method}

Most of the numerical simulations on catastrophic or near catastrophic
collisions have focused on the asteroid belt \citep[for
example,][]{Benz99,Melosh97} and the special case of the formation of
Pluto and Charon \citep{Canup05}. As a result, we will begin our study
by comparing the results from our hybridized method to the
catastrophic collisions of asteroid-like rocky bodies before
investigating more complex icy physical structures.

In high velocity collisions, the shock deformation and the gravitational
reaccumulation phase have very different dynamical times. As a result,
it is difficult to model the entire problem with one numerical method.
In the first phase, the time scale for the propagation of the shock
wave is determined by the target size divided by the sound speed of
the material (few to 10s of seconds for a nonporous rocky target with $R_{T} < 50$ km).  In the
second phase, the time scale for gravitational reaccumulation is
proportional to $1/\sqrt{G\rho}$, where G is the gravitational
constant and $\rho$ is the bulk density (hours for $\rho = 1$ to 3 g
cm$^{-3}$). In order to span the different time scales in our
investigation of catastrophic disruption, we use a hybridized
method that combines a hydrocode to model the shock deformation and an
$N$-body gravity code to model the gravitational reaccumulation.

Hybridized numerical methods have been used by several groups to
investigate the outcome of high velocity impacts in the asteroid belt
over very long time scales.
\citet{Michel01,Michel02,Michel03,Michel04} and \citet{Nesvorny06}
used a smoothed particle hydrocode (SPH) coupled with an $N$-body
gravity code to model asteroid family-forming events. \citet{Durda04}
used the method to investigate binary asteroid formation.  All of
these groups used effectively identical numerical methods which
includes using the same SPH \citep{Benz95} and $N$-body codes
\citep{Richardson00}. The SPH code is used to model the impact, the
propagation of the shock wave, damage accumulation (a proxy for
fracturing), and phase changes. Once the shock wave had decayed and
damage accumulation was complete, the simulation was handed off to the
$N$-body component to model the gravitational reaccumulation phase by
directly translating each SPH particle into an $N$-body particle.

In this study, we model the shock deformation with an Eulerian shock
physics code, \texttt{CTH} \citep{McGlaun90}, instead of the
Lagrangian SPH code used in previous work. Several features of the
\texttt{CTH} code are preferred for an investigation of outer solar
system objects: an adaptive mesh optimizes the resolution of the
problem, mixed cell thermodynamics allows for multiple materials in
each computational cell, and pore compaction models account for the
thermodynamics of high porosity materials. The gravitational
reaccumulation phase of the collision is modeled using the $N$-body
gravity code, \texttt{pkdgrav}
\citep{Richardson00,Leinhardt00,Stadel01}, the same code used in
previous hybrid studies. The combination of a shock physics code and
an $N$-body code allow detailed modeling of the impact as well as
gravitational reaccumulation. The numerical technique is summarized in
the example calculation shown in Fig.~\ref{fig:strip}.

\subsection{CTH}\label{sec:cthmeth}

\texttt{CTH} is a widely-used multi-dimensional and multi-material
shock wave physics software package developed by Sandia National
Laboratories \citep{McGlaun90}. A two-step solution is implemented to
move material through an Eulerian mesh. First, a Lagrangian step
allows the computational cells to distort to follow material motion.
During the Lagrangian step, finite volume approximations of the
conservation equations (mass, energy, and momentum), the material
equations of state, and the constitutive equations are solved
simultaneously. The velocities, energies, stresses and strains are
updated at the end of this phase. Second, a remesh step maps the
distorted cells back to the Eulerian mesh and a second order
convection scheme is used to flux all quantities between cells.  In
this paper, we use two versions of \texttt{CTH} (7.1 and 8.0); both
versions produce similar results. Simulations were carried out in
either 2D cylindrical symmetry or 3D rectangular geometry.

\subsubsection{Adaptive Mesh}

In order to increase the efficiency of the code, \texttt{CTH} also
includes an adaptive mesh refinement (AMR) feature that changes the
computational mesh at user-specified times and locations according to
user-defined criteria \citep{Crawford99}. For example, the primary
criteria are designed to increase the mesh resolution at the moving
locations of (i) free surfaces and material interfaces and (ii) steep pressure gradients. The mesh is refined using an adaptive number of
blocks in the problem domain. Each block is comprised of a fixed
number of computational cells. In this work, we use 10 cells in each
dimension for 2D and 8 cells in each dimension for 3D simulations. In
all simulations, the smallest block size is determined by requiring at
least 20 cells across the radius of the projectile in order to
sufficiently resolve the peak shock pressures \citep{Pierazzo97}.
Refinement subdivides the length scales of a block by factors of 2.
Either 2 or 3 levels of refinement are used in each simulation
depending on the differences in scale between the zeroth level block
size and the smallest block size. This method allows the resolution to
be focused in the most dynamic area of the collision and at material
interfaces (including void spaces). Fig.~\ref{fig:strip}, row one,
shows the evolution of the adaptive mesh for an impact between two
basalt spheres at 1.8 km s$^{-1}$. In this example, the mesh is
refined in front of the shock wave, with moving free surfaces, and
over the entire surface of the projectile.  Note that each box drawn
in the figure represents one block.

\subsubsection{Equations of State}

In this work, the targets and projectiles are composed of either
nonporous basalt or water ice. In both cases, we use a SESAME-style
tabular equation of state (EOS) \citep{Holian84}. In a SESAME table,
the internal energy and pressure are tabulated over a density and
temperature grid.  The tabular format has the advantage of explicitly
representing complex phase diagrams that cannot be captured with
standard analytic models and the disadvantages of being limited by the
finite grid resolution and the developer-defined range of
applicability (e.g., a table may only represent the EOS for a single
phase).  The basalt table is a density-scaled version ($\rho_0 = 2.82$
g cm$^{-3}$) of the SESAME table for $\alpha$-quartz, which includes
the phase transformation to stishovite, melting and vaporization with
dissociation. The table is part of the \texttt{CTH} package and the
development of the table is described in \citet{Kerley99}. For
simulations of impacts into H$_2$O ice, however, pre-existing
equations of state models and tables were inadequate for our needs
(e.g., did not include more than one solid phase or did not accurately
represent the melting curve). Hence, we developed a new SESAME table
for H$_2$O that includes vapor, liquid, and three solid phases (ices
Ih, VI, and VII) that is based on experimentally determined phase
boundaries, shock Hugoniots, and thermodynamic properties
\citep{Wagner02,Frank04,Stewart05,Feistel06}.  For a full description
of the 5-phase H$_2$O EOS, see \citet{Senft08}.  Because the thermodynamic
properties of H$_2$O are difficult to represent with standard analytic
formulations (such as Tillotson \citep{Tillotson62} and ANEOS
\citep{Thompson72}), we decided to use published equations of state
for the individual vapor, liquid and solid phases to build an equation
of state table that is applicable to impact events over the entire
solar system. Note that our H$_2$O equation of state will produce much
more reliable results compared to previous work using the Tillotson or
ANEOS equation of state because it includes the crucial solid phase
transformations to ices VI and VII that control the criteria for
melting ice with a shock wave \citep{Stewart05}.

\subsubsection{Strength Parameters}

The failure of brittle materials (e.g., rocks and ice) is well
described by a pressure-dependent yield surface
\citep[e.g.,][]{Jeager79}, where the shear strength has a value of
$Y_0$ at zero pressure and increases to a maximum value of $Y_M$ with
increasing pressure with a slope defined by the coefficient of
internal friction $\phi_i$. Fragmented materials have lower
zero-pressure shear strength (cohesion, $Y_c$) and friction
coefficient ($\phi_d$) compared to intact materials and the same
limiting $Y_M$.

In all except the hydrodynamic simulations, we use a
pressure-dependent shear strength model, either the geological yield
surface \citep[GEO model,][]{Crawford07} or the ROCK model
\citep{Senft07}.  The yield surface describes the amount of shear stress
that may be supported by the material as a function of ambient
pressure. When a measure of the shear stress (the root of the
second invariant of the deviatoric stress tensor, $\sqrt{J_2}$) is
greater than the shear strength, the material is failing in shear and
undergoes irreversible plastic deformation. At lower stresses, the
material responds elastically. Note that under planar shock wave
conditions with uniaxial strain, the maximum shear stress, $(\sigma_1
- \sigma_3)/ 2$ (where $\sigma$ is a principal stress component), is
equal to $\sqrt{\frac{2}{3} J_2}$.

In the ROCK model, shear strength is linearly degraded from an initial
intact curve to a fragmented curve (friction law) using a dimensionless scalar
measure of damage.  Damage represents the fraction of integrated effective 
plastic strain to failure, which is a proxy for the degree of
fragmentation; hence, intact rock has zero damage and completely
fragmented rock has a damage of one.  The effects of damage are
neglected in the GEO model.  In both strength models, the shear
strength is degraded to zero as the temperature approaches the melting
temperature ($T_m$).

When negative stresses exceed the fracture (tensile) strength, the
material fails in tension and cracks. In \texttt{CTH}, tensile failure
is modeled by inserting void space into a computational cell until the
volume-averaged stress equals the tensile strength, $Y_T$. In the ROCK
model, damage also accumulates with tensile fracturing via a crack
propagation model. Note that both shear and tensile failure lead to
fracturing in brittle materials.

Because of the large uncertainty in the physical properties of
planetesimals, we consider a wide range of strength parameters for
materials with no significant porosity (note that natural intact rocks
generally contain several percent porosity)
(Table~\ref{tab:strength}). A large range of physical
properties is expected based on the inferred range of bulk densities
of outer solar system bodies \citep{Noll08}.  The parameters chosen
are meant to represent general classes of materials, rather than a
specific rock type. For example, a monolithic body may have similar
bulk density and average EOS as a densely packed breccia, but the two
would have very different shear and tensile strengths. A monolithic
rock target is represented by the strong strength parameters. A strong
model with high tension test case is also considered (strongHT). A
breccia with a weak matrix is represented by the weak strength
parameters. An intermediate case (medium) is considered to understand
the systematic dependence on strength parameters and is not meant to
correlate with a particular material group.  Finally, as an end
member, we consider a hydrodynamic (no shear strength) material (e.g., a liquid body) with
tensile strength.

The ROCK model parameters are derived directly from laboratory
quasi-static and dynamic strength measurements (see \citet{Senft07}
for basalt and \citet{Senft08} for ice).  The ROCK model successfully
reproduces crater sizes and damage/fracture patterns observed in
laboratory impact cratering experiments in strong rock
\citep{Senft07}.  The dynamic shear strength of a material is known as
the Hugoniot Elastic Limit (HEL), which constrains the value of $Y_M$.
The HEL of strong rocks is typically a few GPa and the coefficient of
friction is about one \citep{Melosh89}.  The strong model parameters
are based on intact basalt data and are similar to values for other
strong rocks such as granite \citep[see discussion and references
in][]{Senft07,Collins04}.  For comparison, the basalt targets in
\citet{Benz99} were modeled with a Von Mises shear strength (no
pressure dependence) of $Y_M=3.5$~GPa. The weak model parameters fall
in the range observed for terrestrial breccias and cements, with HEL
values between negligible (unmeasurable) to $<1$~GPa
\citep{Willmott07,Tsembelis00,Tsembelis02,Tsembelis04}.  Note that the
HEL of aggregates with a weak matrix is controlled by the strength of
the matrix \citep{Tsembelis06}.  The weak model parameters are
slightly lower than quasi-static data on cold ice and comparable to
warm ice \citep{Senft08}; the temperature-dependent HEL of H$_2$O ice
ranges between 0.05 and 0.6~GPa \citep{Stewart05}.  The ice targets in
\citet{Benz99} were modeled with a constant shear strength of
$Y_M=1$~GPa, much larger than found experimentally \citep[$Y_M \sim
0.1$~GPa,][]{Durham1983} (see also \S~\ref{sec:disc}).

Tensile strength has a strong scale-dependence due to absolute length
(larger bodies have longer fractures; longer fractures are weaker) and
strain rate (impacts by larger bodies have lower strain rates; brittle
materials fail in tension at lower stresses at lower strain rates). As
a result, the tensile strength decreases with increasing size.  We
extrapolate laboratory-scale values for tensile strength to 1-10
km-sized bodies following the method of \citet{Housen90}, where $Y_T
\sim R_T^{-(0.25-0.33)}$ \citep{Housenb99,Ryan98}.  Note that the
uncertainty in the slope of the size dependence leads to an order of
magnitude range of values for tensile strength at km scales (and a
larger range for larger sizes) \citep[see][]{Housenb99}. We use values for the tensile
strength that falls within the predicted range for weak and strong
materials.  Typical laboratory-scale dynamic tensile strengths for
strong rock are order $10^8$~Pa \citep[e.g.,][]{Cohn1981} and decrease
to order $10^6$ to $10^7$~Pa for 1-10 km sized bodies. The strongHT
model, with a laboratory-scale tensile strength, tests the sensitivity
of the disruption calculations to variations in tensile strength.  The
tensile strength of ice is about an order of magnitude smaller than
strong rock \citep{Lange83}.

In the ROCK model simulations, damage initially accumulates primarily
via shear failure during the propagation of the primary shock wave. At
later times, upon reflection of the shock wave from free surfaces,
damage also accumulates in tension. The total damage variable is the sum of the shear and tensile damage \citep{Senft07}. In all cases except for the
smallest targets with the highest tensile strength (strongHT model),
the largest intact fragment is smaller than a computational block
\citep[\S \ref{sec:methval}, see also][]{Benz99,Ryan98}. Hence, each
computational block represents a size distribution of fragments on
similar trajectories.

Although the ROCK model is a more accurate description of the strength
of rocks, the GEO model, with far fewer parameters (6 vs.\ 13), is
easier to intepret. In the GEO model, all simulations use a friction
coefficient of 1, Poisson's ratio of 0.3, and melting temperature of
1474~K.  Simulations using the ROCK model are labeled as such
Tables~\ref{tab:strength} \& \ref{tab:data}, otherwise the GEO model
is used. As described below, both material models produce
quantitatively similar results for gravity-regime catastrophic
disruption calculations.

\subsubsection{Self-gravity}
\texttt{CTH} also has the capability to include self-gravitational
forces in the calculation. The parallel tree method of
\citet{Barnes86} is included in the version 8 distribution.  In
previous work where the intact fragment size distribution is derived,
self gravity is important during the shock deformation phase for
bodies with radii greater than several 10's km, when central
lithostatic pressure becomes comparable to the tensile strength. For
calculations that do not include self-gravity, the model tensile
strength must be appropriately increased to mimic the effect of
lithostatic pressure \citep{Ryan98,Benz99}.  In the simulations
presented here, self-gravity has a negligible effect on the velocity
evolution during the short \texttt{CTH} portion of the calculation.
Hence, it is not utilized for most of the calculations because of the
computational expense. The insensitivity of the results to
self-gravity in the first few seconds is verified by a few test cases
including self-gravity simulations (see \S~\ref{sec:checks} and sims.\ 
19b and 19d)).

\subsection{pkdgrav}\label{sec:pkd}

Once the initial shock wave and the subsequent rarefaction waves have
passed through the target, it is no longer necessary or practical to
continue the simulation with \texttt{CTH}. Almost all of the physics
can be captured with an $N$-body gravity code. In the example shown in
Fig.~\ref{fig:strip}, row 2 shows the pressure gradient in grey scale
versus time. By 60 seconds most of the shock-induced deformation is
complete and gravity is the dominant force.

In these simulations, we have employed the parallelized, Lagrangian,
hierarchical tree code \texttt{pkdgrav}
\citep{Richardson00,Leinhardt00} to complete the gravitational
reaccumulation phase of the collision. \texttt{Pkdgrav} is a second
order symplectic integrator using the leap-frog integration scheme.
The gravitational evolution of particles is modeled under the
constraints of self-gravity and physical collisions. The particles
have a mass and radius, but no explicit equation of state. Material
properties are captured by the coefficient of restitution.

The last output of \texttt{CTH} is run through a translator to convert
the Eulerian grid data into Lagrangian particles, creating initial
conditions for \texttt{pkdgrav} (Fig.~\ref{fig:strip} rows 3 and 4).
If the \texttt{CTH} calculation is in 2D, the last time step is
rotated about the axis of symmetry creating a 3D object. Each adaptive
mesh block is then an annulus of material. The annuli are divided into
spherical particles with a radius determined by the smallest of the 2D
block dimensions. If the \texttt{CTH} calculation is 3D, the adaptive
mesh blocks are translated into spherical particles with a radius
determined by the smallest dimension of the block. To eliminate
simultaneous collisions, which \texttt{pkdgrav} cannot tolerate, a
small amount of random noise is added to the velocity of each particle
($1\%$ the escape speed) when handing off from \texttt{CTH}. Because
each \texttt{CTH} block represents a size distribution of particles,
no provision is made to bind certain \texttt{pkdgrav} particles together as an
intact fragment, although this would be necessary to model the
catastrophic disruption of bodies of smaller bodies that transition
from the gravity to the strength-dominated regime.

If there is only a small fragment of material within a block, a
straight translation into a Lagrangian particle would create a very
low density particle with an artificially large radius. Therefore, to
avoid unrealistic particles, the minimum density of each particle is
limited to 0.75 of the original bulk density of the target. If the
mass of a block is less than $4\times10^{-7} M_{\rm tot}$, no particle
is created. The cutoff value is an empirical threshold that eliminates
numerous tiny ejecta particles that do not accumulate into the largest
remnants.  In addition, little ejected material leaves the
\texttt{CTH} mesh before handoff. Therefore, the total amount of mass
discarded in the translation is negligible and usually involves
material that is escaping from the largest remnant of the collision
event.

To facilitate translation into spherical particles, the geometry of
the adaptive mesh blocks are kept close to square or cubical. The
total mass of the square/cubic block is conserved upon translation,
thus the mass density of the \texttt{pkdgrav} particle is about a
factor of two higher than the mass density in the original
\texttt{CTH} block. This means that some assumption must be made about
the macroporosity of the post-collision remnants in order to determine
a radius from the mass of the remnant. 
Note that in the work presented here, negligible amounts of vaporized
material is produced.  Thus, the effects of volume changes are ignored
in the translation.

Each \texttt{pkdgrav} particle is modeled as an indestructible sphere.
Collisions between particles follow one of three scenarios: perfect
merging, conditional merging, or inelastic bouncing.  Tangential and
normal coefficients of restitution are used to govern the amount of
kinetic energy lost as a result of each particle collision
\citep{Richardson94,Leinhardt00}. The impact velocity is given by
${\bf v}={\bf v}_{n}+{\bf v}_{t}$, where ${\bf v}_{n}$ is the
component of the velocity vector that is normal to the surface of the
target particle and ${\bf v}_{t}$ is the tangential component. The
post-impact velocity is given by ${\bf v}'=-\epsilon_{n}{\bf
  v}_{n}+\epsilon_{t}{\bf v}_{t}$, where $\epsilon_{n}$ is the normal
coefficient of restitution and $\epsilon_{t}$ is the tangential
coefficient of restitution. Both $\epsilon_n$ and $\epsilon_t$ have
values between 0 and 1. In this paper, perfect merging ($\epsilon_n =
0$) is used in most cases to reduce calculation time (denoted by
`merge' in the simulation type in Table~\ref{tab:data}); however,
information is lost about the shape and structure of the reaccumulated
body. In order to examine the structure of a remnant, inelastic
collisions ($\epsilon_n$ is a nonzero constant based on the type of
material) are used in some 3D cases (see \S~\ref{sec:proplr}).
Conditional merging ($\epsilon_n = 0$, if the collision speed is less
than the escape speed, otherwise $\epsilon_n > 0$) was used to verify
the results of simulations using perfect merging.  Recent field
observations and friction experiments on rocky materials constrain the
value of $\epsilon_n$ \citep{Chau02}.  In general, $\epsilon_n$ varies
between 0.1 and 0.5 for rocks, soils, and ice
\citep{Chau02,Higa96,Higa98}. In this work, simulations denoted by
`bounce' in Table~\ref{tab:data} use $\epsilon_n=0.5$ and simulations
denoted by `loweps' use $\epsilon_n=0.1$. In all cases, $\epsilon_t$
is fixed equal to 1, which implies negligible surface friction as
found in experiments on rocky materials \citep{Chau02}.

The \texttt{pkdgrav} phase of the calculation continues until the
largest post-collision remnant reaches dynamical equilibrium.
Dynamical equilibrium is defined by the criteria that the total mass
accreting and orbiting the largest post-collision remnant must be
$<10\%$ of the mass of the largest remnant ($M_{lr}$); in
addition, a minimum of 50 dynamical times is imposed upon all
calculations.

In this work, the number of \texttt{pkdgrav} particles ranges from
several thousand to several tens of thousands ($N_{init}$ in
Table~\ref{tab:data}). Recall that each \texttt{pkdgrav} particle
represents a \texttt{CTH} computational block with $8^3$ cells in 3D
and $10^2$ cells in 2D. The AMR feature allows the shock deformation
to be very well resolved while minimizing the number of particles in
the \texttt{pkdgrav} phase of the calculation.  Hence, the
\texttt{CTH-pkdgrav} technique highly resolves the energy and momentum
coupling to the target at relatively low computational expense.

\subsection{Tracers}

To track the motion and physical properties of material flowing
through the mesh, massless Lagrangian tracer particles are included in
the \texttt{CTH} calculation to provide material histories, including
trajectories, pressures, temperatures, etc.  In the 2D \texttt{CTH}
calculations, 900 tracer particles are placed in the target on rays
initiating from the impact site (Fig.~\ref{fig:2dtrace}A).  In the 3D
\texttt{CTH} calculations, tracer particles are placed in a uniform 3D
grid in both the projectile and target. At handoff from the hydrocode
to the gravity code, the tracers associated with each \texttt{pkdgrav}
particle are recorded for analyses of the material deformation history.
In Fig.~\ref{fig:2dtrace}B, particles that contained no tracer
particles are assigned a peak pressure history value by averaging the
peak pressure from the nearest 2 neighbors.

It is difficult to ensure that the divergent flow in the ejecta cone
is filled with tracer particles. Because it is highly resolved and
comprises a significant surface area, the ejecta cone requires the
most interpolation.  However, most of the material in the ejecta cone
escapes after the impact. As a result, interpolation of a large amount
of the ejecta cone will not effect any conclusions that are drawn
about the properties of the largest remnant (\S \ref{sec:proplr}).
The original surface of the target near the impact site also needs to
be interpolated because it has been highly resolved by the
adaptive mesh refinement, resulting in \texttt{pkdgrav} particles
between tracer particles.

\subsection{Handoff Time}

In the hybridized method presented here, the transition from hydrocode
to $N$-body gravity code occurs after the shock and rarefaction wave
has propagated through the system and dissipated. After this point,
modification of material due to shock waves is negligible.  Figure
\ref{fig:handoff} shows the mass of the largest post-collision remnant
versus time of handoff from \texttt{CTH} to \texttt{pkdgrav} for an
impact between a 10 km radius target and a 0.83 km radius projectile
at 1.9 km s$^{-1}$ (sim 8c in Table \ref{tab:data}). The mass of the
largest remnant levels off between 10 and 15 seconds or about four to
six times the sound crossing time. A least squares fit of the data
after 10 seconds finds a slope consistent with zero ($-0.002\pm0.003$).
There is scatter in the mass of the largest remnant due to varying
particle number and particle size from simulation to simulation. The
resolution of the $N$-body component of the simulation varies
depending on the handoff time since \texttt{CTH} utilized adaptive
mesh refinement. We estimate a conservative error in the mass of the
largest remnant to be $\pm 0.1$ $M_{lr}/M_{T}$, where $M_T$ is the
target mass, based on the peak to peak scatter of the test results. In
the rest of the simulations presented in this paper, the handoff time
was $>$ 4 times the sound crossing time of the target.

\section{Results}\label{sec:results}

We conducted sets of simulations with (i) fixed projectile-target size
ratios and varying impact velocity or (ii) fixed impact velocity and
varying projectile size. In this work, we are primarily concerned with
the mass and physical properties of the largest remnant. The critical
disruption energy, $Q_D^*$, is determined by a non-linear least
squares fit to the resolved largest remnants ($M_{lr}/M_{T} > 0.1$).
Table~\ref{tab:data} presents a summary of the simulation parameters
and results.  The simulations denoted with an asterisk are close to or
below the resolution limit of the hybrid calculations ($0.1$
$M_{lr}/M_{T}$) and have a low number of \texttt{pkdgrav} particles
($N_{lr}<200$). (If it is necessary to use an under-resolved
simulation to determine $Q_D^*$ the under-resolved simulation is given
less weight than the resolved simulations). Tables \ref{tab:qstar} \&
\ref{tab:qstarcv} present the critical impact velocity (for a fixed
projectile to target mass ratio) and critical projectile radius (for a
fixed impact velocity) and the corresponding $Q_D^*$ values for each
set of simulations (Fig.~\ref{fig:qstar}). We vary the strength,
impact angle, \texttt{pkdgrav} collision model, and composition to
determine their effects on the energy coupling in catastophic
disruption events and the physical properties of the largest
reaccumulated remnants.

Next, we describe the validation of the hybrid method by comparison to
previous work (\S~\ref{sec:methval}).  We conduct simulations with
parameters comparable to results on the catastrophic disruption of
strong bodies presented in \citet{Melosh97} (sims.\ 2, 10, \& 16) and
\citet{Benz99} (sims.\ 7 \& 12). In section \S~\ref{sec:catdis}, we
describe the results of varying the shear and tensile strength of the
bodies while holding the target to projectile size ratio fixed
(Table~\ref{tab:strength}). In 3D simulations, we consider the effects
of impact angle and composition on the physical properties of the
largest remnant (\S~\ref{sec:proplr}). Finally, we examine the role of
the \texttt{pkdgrav} collision model on the gravitational
reaccumulation process (\S~\ref{sec:coefrest}).

\subsection{Method Validation}\label{sec:methval}
The catastrophic disruption criteria derived from our hybrid
calculations are presented in Figure~\ref{fig:qstar} and
Tables~\ref{tab:qstar} \& \ref{tab:qstarcv}).  In
Figure~\ref{fig:qstar}, the results are compared to the $90^{\circ}$,
strong target simulations by \citet{Benz99} (solid lines) and
\citet{Melosh97} (dashed lines). Although the details of the shear and
tensile strength models are different, the strong target results
(green and red symbols) are in good agreement with previous work.
The weak and hydrodynamic targets (black and blue symbols) have
disruption criteria that are significantly smaller than the strong
target cases (see \S~\ref{sec:catdis}).

In this work, we held the size ratio between the target and the
projectile fixed for most simulations rather than holding the impact
velocity constant in order to ensure $M_{P} \ll M_{T}$. With the fixed
target and projectile sizes, the optimum resolution for the
calculations can be easily controlled. In previous work with a fixed
impact velocity \citep[e.g.,][]{Benz99}, the size and resolution of
the projectile varied significantly (and in some cases $M_P > M_T$).
In simulations 2, 10, and 16 (red points in Fig.~\ref{fig:qstar}), the
size ratio between the projectile and targets are the same as in
\citet{Melosh97}. The derived critical impact velocities (that
correspond to the fitted $Q^*_D$ in Table~\ref{tab:qstar}) are 1.4,
4.8, and 1.9~km~s$^{-1}$, respectively, compared to 1.3, 3.7, and
1.8~km~s$^{-1}$ in \citet{Melosh97}. Although our results are
systematically higher, perhaps indicating that our shear strength
parameters are slightly stronger than the unreported values used by
\citet{Melosh97}, the critical impact velocities are generally within
our error bars. Simulations 7 and 12 have fixed impact velocities of
3~km~s$^{-1}$, as in \citet{Benz99}, and our results agree within
error (open green squares and solid line in Fig.~\ref{fig:qstar}).
Note that there is comparable scatter in the individual $Q_D^*$ data
points that are used to derive $Q^*_D$ curves between our and previous
work.

Another measure of the accuracy of our hybrid simulations is the mass
of the largest remnant as a function of impact energy. \citet{Benz99}
found that the normalized mass of the largest remnant, $M_{lr}/M_{T}$,
follows a tight linear trend with the normalized impact energy,
$Q/Q_D^*$, in the hypervelocity regime. For example, they fit a slope
of $-0.5$ for 3~km~s$^{-1}$ impacts into basalt targets. We find a
similar slope of $-0.48\pm0.02$ (Fig.~\ref{fig:mlrvsQ}) for impact
velocities over the larger range of 0.7 to 5.6~km~s$^{-1}$ into basalt
targets.  The linearity of the results justify the use of a linear fit
to the remnants from each simulation within a group to derive the
value of $Q_D^*$.  Note that the slope would be -1 for pure energy
scaling (Stewart \& Leinhardt, in prep.).  The slope and linearity of
the largest remnant mass is a satisfying validation of the hybrid
technique. The tight linear relationship is independent of the impact
velocity and the strength of the targets, which varies from
hydrodynamic to strong.  \citet{Benz99} noted a slight dependence of
the linear slope on impact velocity, which may include differences in
the projectile to target size ratio. \citet{Benz99} also observed
slightly steeper linear slopes for the normalized mass of the largest
remnants in ice targets.

Hence, we confirm that the \texttt{CTH-pkgrav} hybrid technique
is in excellent agreement with previous studies of catastrophic
disruption in the gravity regime.

\subsubsection{Comparing Different Impact Velocities}
It is well established that changing either the impact velocity or the
mass ratio between the projectile and target will result in
differences in the energy and momentum coupling to the target
\citep[e.g.,][]{Housen90,Melosh97}.  Therefore, detailed comparison of
catastrophic disruption results between different studies and between
different target sizes requires an explicit correction for the
different modeled impact velocities and/or mass ratio.  In order to be
able to extend our results to different (hypervelocity) impact
velocities (where $M_P \ll M_T$), we apply a impact velocity
correction using the fragmentation theory developed by
\citet{Housen90}.

First, we use the slope of the $Q_D^*$ curve to constrain a material
parameter, $\mu$, in the coupling theory of \citet{Housen90}.  In the
gravity regime, $Q_D^*$ would be proportional to $R_{T}^2$ if material
properties were inconsequential and pure energy scaling applied. In
reality, the coupling of the projectile's kinetic energy to the target
depends on the material and the impact velocity. In the analytical
model of \citet[][Eq.~36]{Housen90} for the critical shattering criteria,
$Q_S^*$, material properties are captured by a coupling parameter that
scales with V$_i^{\mu}$ rather than scaling with the impact energy.
The exponent, $\mu$, is a simple descriptor of material properties
with a value of 0.55 to 0.60 for rocks and water and $\sim 0.4$ for
sand \citep{Housen90,Holsapple93,Schmidt87}. In the gravity regime,
$Q_S^*$ $\sim R_{T}^{3\mu}$ $\sim R_{T}^{1.2-1.8}$ for sand to rocky
targets \citep[][Eq.~68]{Housen90}.

Next, we assume that the additional energy required for dispersal of
the shattered fragments produces a negligble change in slope between
the $Q_D^*$ and $Q_S^*$ curves. This assumption is supported by
\citet[][Eq.~73]{Housen90}'s estimate of $Q_D^*$ based on the fraction
of shattered fragments that reach escape velocity. Using laboratory
data as a constraint on the fragment velocity disruption, the exponent
of the $Q_D^*$ curve is negligibly smaller than the $Q_S^*$ curve in
the gravity regime.  Numerical simulations of catastrophic disruption
support the departure from energy scaling and similarity in slope to
$Q_S^*$. The 90$^{\circ}$, basalt, gravity regime data from
\citet{Benz99} ($V_i=$3 km s$^{-1}$) and \citet{Melosh97}
($V_i=$1.3-5.3 km s$^{-1}$) have fitted exponents of 1.26 and 1.44,
respectively, implying that the value of $\mu$ falls in the range of
0.42 to 0.48.

Laboratory and numerical experiments demonstrate that increasing the
impact velocity increases the critical disruption energy for a fixed
target radius \citep{Housen90,Benz99}.  With increasing hypervelocity
impact velocities, more of the impact energy is expended in material
deformation and the disruption process becomes less efficient.
\citet[][Eq.~68]{Housen90} find that $Q_S^*$ scales by $V_i^{-3\mu+2}$
in the gravity regime, and we assume that the same velocity correction
applies to $Q_D^*$. The derived correction factor is in reasonable
agreement with results from \citet{Benz99}, $Q_D^*$ increases by a
factor of $\sim$1.7 between lines of constant $V_i=3$~km~s$^{-1}$ and
$V_i=5$~km~s$^{-1}$ for basalt targets. In addition, in our own
simulations, the velocity corrected fixed mass ratio simulations 6 \&
11 are consistent with the constant impact speed varying mass ratio
simulations of the same strength model 7 \& 12. The original critical impact velocities (pre-correction)
are noted in Table~\ref{tab:qstar}.

We illustrate the velocity correction in Figure~\ref{fig:qstar}B,
where the $Q_D^*$ data are adjusted to a constant impact velocity of 3
km s$^{-1}$. The correction uses an empirical value of $\mu \sim 0.45$
based on a mean gravity regime slope ($3\mu \sim 1.35$) from our and
previous worker's results.  Our hybrid simulations of strong and weak
basalt targets are fit with slopes of 1.3 and 1.2 respectively (dotted
lines in Fig.~\ref{fig:qstar}B).  Interestingly, the two hydrodynamic
cases suggest a slope of 1.9, approaching the ideal $Q_D^*\sim
R_{T}^2$ for pure energy scaling in the gravity regime.

Note that the dependence of $Q^*_D$ on shear strength is confirmed to
be a primary result and not an artifact of the differences in critical
impact velocities. The physical explanation for the role of shear
strength is dicussed in \S~\ref{sec:catdis}.

\subsubsection{Other Checks \label{sec:checks}}
We confirm that each \texttt{pkdgrav} particle represents a size
distribution of smaller fragments. The targets and projectiles are
completely damaged ($D=1$) for all simulations except the strongHT
cases with $R_T=2$~km. In the strongHT case, the tensile strength is
the highest expected for km-scale bodies. These results are consistent
previous work, where the mass fraction of the largest intact fragment
in the reaccumulated remnant drops precipitously above target radii of
a few 100~m \citep{Benz99,Melosh97}. In the 2-km strongHT cases
(sims.\ 2, 6R, 7R), the largest fragment is composed of several
\texttt{pkdgrav} particles that all merge into the largest
gravitationally reaccumulated remnant.  Some of the the scatter in the
$R_T=2$~km results reflects the increased influence of shear and
tensile strengths compared to the larger bodies.  However, the
amplitude of the strength effects at this size are within error within
each strength model group.

In simulations 1-16, we use 2D geometry in \texttt{CTH} to decrease
the computation time.  We verify that calculations using 2D geometry
are robust by comparing identical impact conditions in 2D and 3D
geometries. The mass of the largest remnant in simulation 17a
($M_{lr}/M_{T}=0.57$) is nearly identical to the mass of the largest
remnant in simulation 14c ($M_{lr}/M_{T}=0.60$). Therefore, we have
determined that simulations using 2D and 3D \texttt{CTH} components
are in very good agreement with one another.

Finally, we confirmed that self-gravitational forces are negligible
during the \texttt{CTH} component of the calculation. Simulations 19b
and d include self-gravity in the \texttt{CTH} component. The mass of
the largest remnant and the second largest remnant is virtually
unchanged when compared to simulations without self gravity in the
\texttt{CTH} component (sim 19a and c).  In addition, the resolution
in simulations 19c and d are higher than in simulations 19a and b.
Once again, the masses of the largest post-collision remnants are
indistiguishable.

\subsection{Strength Effects on $Q_D^*$}\label{sec:catdis}

Small bodies in the solar system possess a wide range of shear
strengths, from monolithic bodies inferred from fast spin rates, \citep[e.g.,][]{Pravec2000} to
nearly strengthless bodies \citep[e.g., $\sim 1-10$~kPa for comet
9P/Tempel 1,][]{Richardson07}. Previous simulations have shown that
the catastrophic shattering criteria ($Q_S^*$) is typically two orders
of magnitude less than the criteria for dispersal ($Q_D^*$) in the
gravity regime \citep{Melosh97}. As a result, the role of strength has
largely been ignored in the gravity region of the $Q_D^*$ curve.
Although previous workers have considered a wide range of materials,
including ice, basalt and mortar, the literature has focused on the
differences in tensile strength and not considered a wide range of
shear strength \citep{Ryan98,Benz99}.

In this work, we consider a wide range of shear and tensile strengths.
We find that the catastrophic disruption criteria is higher in targets
with higher shear strength. As expected, the effect of tensile
strength is negligible (Fig.~\ref{fig:qstar}). An increase in the
tensile strength by two orders of magnitude (red symbols: $Y_T = 10^5$
Pa, green symbols: $Y_T = 10^7$ Pa) has a small effect on the
collision outcome at a target radius of 2 km (within error) and
essentially no effect on the collision outcome at a target radius of
10~km. In contrast, the shear strength model has a large effect on the
resulting $Q^*_D$, with variations up to a factor of 8 at $R_T=10$~km
between weak and strong targets.

The shear strength affects the catastrophic disruption criteria in the
gravity regime by modifying the pressure and velocity distributions in
the target. Shear strength increases the peak shock pressure and shock
wave decay rate compared to a strengthless material. In
Fig.~\ref{fig:strength}A, the peak shock pressures along a ray of
tracer particles adjacent to the centerline (refer to
Fig.~\ref{fig:2dtrace}) are shown for identical impact conditions but
varying shear strength (sims.\ 8a, 10c, 11c, and 13a). The peak
pressure is higher with greater shear strength because the shock
Hugoniot is shifted upward by 2/3 of the yield strength compared to a
hydrodynamic Hugoniot \citep{Melosh89}.  However, as the shock wave
propagates into a target with higher shear strength, a portion of the
energy is partitioned into plastic deformation and the shock wave
decays more rapidly compared to a weaker material. In other words, the
stronger the material, the more energy is partitioned into overcoming
the shear strength.  

The cumulative distribution of peak pressures reflects the different
shock decay profiles (Fig.~\ref{fig:strength}B). Because of the
shallower shock decay rate in the weak target, a larger mass fraction
reaches higher shock pressures compared to the strong targets. The
fraction of highly shocked material decreases as the strength of the
material increases. The initial velocity evolution after the impact is
controlled by the shock pressure field. Because the weak target has a
shallower shock decay profile, the pressures are higher at the time
the shock wave encounters the free surface. The material velocity
follows pressure gradients, so the velocity distribution in the weak
target is faster compared to the strong targets
(Fig.~\ref{fig:strength}C). Thus, more material reaches escape
velocities and the largest reaccumulated remnant is smaller for weak
targets.

Note that the pressure profiles using the GEO model are very similar
to the ROCK model (e.g., sims.\ 10c \& 11c in
Fig.~\ref{fig:strength}A), which leads to similar $Q_D^*$ values for
each of the strength groups (see Table~\ref{tab:qstar}: weak -- 1,
4R, strong -- 2, 5R, strongHT -- 3, 6R). As expected, the pressure
decay profile in the medium strength target falls in between the weak
and strong target cases.

Using a physical model derived from impact cratering, \citet{Melosh97}
used the pressure decay profile to predict the catastrophic disruption
criteria, assuming an average power law decay exponent of -2. The
material velocity at the antipode from the impact is derived from the
peak pressure using the conservation equations and assuming velocity
doubling at the free surface.  \citet{Melosh97}'s catastrophic
disruption criteria was assumed to correspond to an antipodal velocity
of half the escape velocity of the target. From
Fig.~\ref{fig:strength}A, it is clear that shear strength has a
significant effect on the pressure decay profile and the resulting
material velocities. The antipodal velocities range from about
20~m~s$^{-1}$ for the strong target cases to 40~m~s$^{-1}$ for the
weak target. Both cases are greater than the escape velocity, even
though the strong target case is subcatastrophic. Although the
conceptual model from \citet{Melosh97} is a very useful approximation
for a disruption criteria, the true velocity distribution is more
complicated and antipodal velocities alone are insufficient to predict
disruption.

We have found that the shear strength affects the catastrophic
disruption criteria by modifying the pressure and velocity
distributions in the target.  In this work, we considered general
groups of material properties. We plan to conduct a more focused
investigation on the specific properties of pure ice and ice-rock
mixtures in a later paper using ROCK model parameters for ice that are
currently under development \citep{Senft08}.

\subsection{Properties of the Largest Remnant}\label{sec:proplr}

Because the gravitational reaccumulation stage is calculated in our
hybrid technique, we are able to examine the physical properties of
the reaccumulated remnants.  Although we are interested in all of the
collision remnants, it is the largest remnant that has the highest
numerical resolution. The physical properties of large collision
remnants may also be observed astronomically. For example, some of the
largest Kuiper belt objects are inferred to be rubble piles
\citep{Trilling06,Lacerda07}.

First, we consider the range of shock deformation experienced by the
material that reaccumulates into the largest remnant.  In
Fig.~\ref{fig:strength}D, the peak shock pressure distribution is
shown for the largest remnant in the four simulations with identical
impact conditions but different strength models.  When compared to the
shock pressure distribution of the total mass
(Fig.~\ref{fig:strength}B), we see that the largest remnant materials
have experienced a large range of shock pressures in the collision
event. In general, as expected, the shock pressures are lower in the
largest remnant, but some fraction of the remnant is derived from the
most highly shocked material. Although peak shock pressures are not
particularly high compared to other hypervelocity impacts in the solar
system, if the material responds in an irreversible way to low shock
pressures (e.g., by structural changes, phase changes, or chemical
changes), some of the materials in the largest remnant will reflect
these changes.

Next, we examine the effect of impact angle on the largest remnant. We
have conducted most of our simulations at 90$^\circ$ for computational
efficiency, but in general, an impact between two small bodies in the
outer solar system will not be head-on. In order to illustrate the
effect of impact angle, we compare two impact events on a 50-km radius
target that have the same mass largest post-collision remnant
($M_{lr}/M_T\sim0.7$). Figures \ref{fig:angeff}A and B show the
cumulative peak pressures attained in a 90$^\circ$ (sim.\ 14d) and a
45$^\circ$ (sim.\ 18a) impact.  The solid line is peak pressure
attained by the total mass, the dashed line is the largest remnant,
the dotted line is the second largest remnant.  Both the largest and
second largest remnants show a similar range in peak pressure
distribution as the total mass. Note, however, that the second largest
remnant is generally under-resolved in our calculations. The finding
that material in the largest remnant experiences the full or nearly
full range of peak shock pressures is a general result from all of our
simulations.  The cumulative pressure distribution is wider, meaning a
larger range of shock pressures, for the 45$^{\circ}$ case compared to
the 90$^{\circ}$ case. The 90$^{\circ}$ impact has a larger mass
fraction of material in the high pressure tail, while the 45$^{\circ}$
impact has a larger mass fraction in the low pressure tail.  Because
an oblique impact has less efficient energy coupling to the target,
the impact velocity must be faster to reach the same outcome (same
$M_{lr}/M_T$).  As a result, some of the material reaches a higher
peak pressure.

We also examine the provenance of material that reaccumulates into the
largest remnant in catastrophic-level collisions, where $M_{lr}/M_T
\sim 0.5$. Previous studies have noted that most material originates
from the antipoldal region to the impact, but not examined the source
and reaccumulation in more detail.  Fig.~\ref{fig:prov} shows the
original (pre-collision) location of particles that reaccumulate into
the largest remnant for 3D simulations of 50-km radius targets (sims.\ 
17-19). Particles are color coded for peak shock pressure. The left
column shows simulations where the perfect merging criteria is used in
the \texttt{pkdgrav} phase. The right column shows the same
simulations with inelastic collisions (bouncing) with a normal
coefficient of restitution of 0.5. The striking differences between
the merging and bouncing simulations are discussed in
\S~\ref{sec:coefrest}. The true provenance of material in the largest
remnant should fall between these two limiting cases.

In all simulations, the largest remnant is primarily composed of
material from the antipodal hemisphere suggesting any compositional
diversity from the original target could be preserved in the largest
remnants in catastrophic-level collisions. Note that the
pressure range in the ice target is lower compared to the basalt
targets. The lower shock pressure range results in a generally lower
material velocity distribution. In the gravity regime, ice targets
disrupt at lower specific impact energies, primarily because of the
lower density of the ice and the resulting smaller escape velocity.

Although the interior structure of small bodies in the outer solar
system cannot be observed directly, their surface properties can be
studied and internal structures can be inferred. The results from our
simulations can help to determine whether features observed on the
surface can be extended to the interior. Fig.~\ref{fig:surf} shows an
example of a largest remnant from a 3D, 45$^{\circ}$ impact into a
50-km radius basalt target (sim.\ 17b).  The exterior of the remnant is
heterogeneous with particles that have a broad range of peak pressures
(colors ranging from yellow to purple). The interior of the largest
remnants is slightly less heterogeneous - composed mostly of
moderately shocked material (mostly green). This is a general result
from all three largest remnants of the inelastic collision (bouncing)
simulations (17, 18, and 19).  

The same result is shown more quantitatively in the cumulative
pressure plots in Fig.~\ref{fig:intvsext}, which presents the results
from inelastic collision simulations of 50-km targets. The interior
materials (dotted line) have a steeper distribution than the exterior
(dot-dashed), meaning that the interior experienced a more confined
range of shock pressure compared to the exterior materials.  The
exterior material experienced both higher and lower shock pressures.
However, although this result suggests that there may be a core of
antipodal material that persists intact during the disruption event,
we find that this is not the case.  The largest remnant is composed of
material that originates from random radii in the target, hence any
radial gradients in the original target are destroyed and not present
in the largest post-collision remnant.

\subsection{Collision Models and Gravitational Reaccumulation}\label{sec:coefrest}

Reaccumulation of remnants after a catastrophic disruption event
involves slow to moderate velocity collisions and gravitational
torques. The cloud of fragments is a mixture of dust, fractured
fragments, competant fragments, and their gravitational aggregates.
Each collision during reaccumulation is an impact calculation in its
own right. This process must be computationally simplified.
\citet{Benz99} and \citet{Melosh97} used analytical methods to
determine the mass of the largest post-collision remnant. In the
simulations presented in this paper the largest remnant is determined
directly from numerical integration in \texttt{pkdgrav} (Fig.
\ref{fig:strip}).

During reaccumulation, the physical processes that govern the outcome
of collisions is assumed to be captured by the coefficient of
restitution. We consider three different scenarios (as mentioned in \S
\ref{sec:pkd}), where the collision outcome is (i) perfectly inelastic
(perfect merging, $\epsilon_n = 0$), (ii) conditional merging (if $V_i
> V_{\rm esc}$ then inelastic bouncing with $\epsilon_n > 0$ else
merging with $\epsilon_n = 0$), and (iii) inelastic bouncing
($\epsilon_n > 0$). In case one, merging replaces the two colliding
particles with a single particle of the combined volume and average
density, conserving angular momentum. Perfect merging is the most
computationally efficient, however, any particle that collides with
another particle will merge regardless of the impact speed. Therefore,
perfect merging may result in the largest post-collision remnant being
unphysically massive. Conditional merging (case two) allows mergers
only when the impact speed is less than the escape speed of the two
particles, otherwise the particles rebound with energy loss determined
by the coefficient of restitution.  This is a more physical model;
however, in both of the merging scenarios, information about the
shape, surface, or internal structure of the remnants are not
preserved.  Inelastic collisions (case three) between particles is
computationally expensive because the number of particles stays
constant over the time of the simulation and the number of collisions
that must be calculated is large \citep{Leinhardt00}. Inelastic
collisions preserves the shape and structural information about the
remnants.

Although all of these models are simplifications of the true
reaccumulation process, perfect merging and inelastic bouncing
represent the extremes, and they bound the true mass of the largest
remnant. In Fig.~\ref{fig:prov}, it seems that increasing the
coefficient of restitution significantly decreases the critical
catastrophic disruption energy. For example, under the same impact
conditions, the largest remnant may be $\sim 3$ times larger using
perfect merging versus inelastic bouncing (e.g., sim.\ 17a vs.~17b).
The most striking difference between the two cases is the amount of
most highly shocked material that reaccumulates into the largest
remnant. 

However, closer examination of the end member cases reveals that the
coefficient of restitution is not the controlling factor in the
outcome of the \texttt{pkdgrav} calculation. In order to confirm the
robustness of the perfect merging case, all simulations were also run
using the conditional merging criteria. The conditional merging
results were nearly identical to the results of the perfect merging
simulations, both in the mass of the largest remnants and the
provenance of the material. We also tested the dependence of our
bouncing simulation results on the value of the normal coefficient of
restitution by running 3D simulations with two different values:
$\epsilon_n = 0.5$ (sim 17b, 18b) and 0.1 (sim 17c, 18c).  The
different values for $\epsilon_n$ had no significant effect on the
outcome of simulations using inelastic collisions. The lack of
dependence on $\epsilon_n$ is due to the nature of the problem, where
the majority of collisions occur between particles that are
gravitationally bound to the largest remnant.

Finally, we examined the criterion for reaching dynamical equilibrium in
the reaccumulation of the largest remnant. As a comparison to the
criteria described in \S~\ref{sec:pkd}, we also used the more
sophisticated algorithm in the \texttt{companion} code
\citep{Leinhardt05}. We calculated the amount of material bound to the
largest remnant using the hierarchical system search feature of
\texttt{companion}. The results are in excellent agreement, as we find
that the mass of the largest remnant does not omit any significant
mass from an extended bound system.

We conclude that the results from the end member models (merging and
bouncing) are computationally robust.  Therefore, the differences in
the merging versus bouncing results are due to the underlying physical
differences in the models and not due to the value of a model specific
parameter.

A significant amount of the difference between the merging and
bouncing results can be explained as the result of runaway merging in
both the perfect and conditional merging models. The first particles merge very
early in the \texttt{pkdgrav} part of the simulation. The merged
particles are replaced with a new particle that conserves volume at
their center of mass. As a result of the modified geometry upon
merging, the new particle will very likely overlap neighboring
particles, which in turn will most likely merge with the new particle.
The process results in an artificial increase in the mass of the
largest remnant. Note that all previous studies using the hybrid
hydrocode-to-$N$-body code technique have used the perfect or
conditional merging criteria and suffer from the same artifact.  

On the other hand, the inelastic bouncing simulations probably
underestimate the mass of the largest remnant. At the time of handoff
between \texttt{CTH} and \texttt{pkdgrav}, particles at the base of
the impact crater begin to bounce when in reality they would continue
to move down into the target and shear (and not bounce at all). The
abrupt transition from shearing flow to discrete, bouncing particles
artificially disrupts a portion of the evolution of the problem. As a
result, we believe that reality lies somewhere in between the merging
and bouncing results. In future work, we will investigate these
technical issues further by examining the modification in the material
flow field from the handoff process.

\section{Discussion} \label{sec:disc}  

Predicting the outcome of collisions in the outer solar system is
challenging because there are several competing effects. Outer solar
system bodies most likely have lower strength in comparison to
asteroids of a similar size, yet they may be very porous which may
help limit the damage from an impact \citep{Asphaug98}. Small outer
solar system bodies should contain a high fraction of volatiles but
this may be mixed with other materials, either intimately or
segregated into layers (differentiated), which could have a
complicated effect on the collision outcome.  We have developed a new
hybrid numerical technique that has the capability of tackling all of
these complexities. In this work, we focus on nonporous bodies of
varying shear strength and composition.

Using our hybrid numerical technique, we derive new catastrophic
disruption criteria for nonporous bodies of varying strength.  Figure
\ref{fig:compqstar} presents a range of $Q_D^*$ curves for outer solar
system bodies, including both the strength and the gravity regimes.
We focus on the mean impact velocity in the present-day Kuiper belt,
about 1~km~s$^{-1}$ \citep{Trujillo01}. The 3~km~s$^{-1}$ disruption
curves for strong basalt and ice from \citet{Benz99} are presented as
upper limits in strength and velocity for the Kuiper belt. As noted by
\citet{Benz99}, their 0.5~km~s$^{-1}$ ice simulations produced results
that were not consistent with the faster simulations and should be
considered with caution until further work is done on the disruption
of icy bodies.

The critical disruption energy is a central component of models of the
collisional evolution of a population of small bodies.  $Q_D^*$ curves
are important because the transition from the strength to gravity
dominated regimes defines the minimum required energy for disruption
and the size of the weakest bodies. Thus, the amount of collisional
grinding during the evolution of the solar system is sensitive to the
disruption criteria \citep[eg.][]{Kenyon04}. In addition, the
transition from strength to gravity produces observable kinks in the
size distribution of small bodies, as found in the asteroid belt
\citep[e.g.,][]{OBrien05}.  

At present, the best $Q_D^*$ curves for the outer solar system are
still inadequate. Small bodies in the outer solar system are diverse,
and we expect a wide range composition, bulk density, and shear
strength. Here, we present our best recommendation for nonporous
bodies.  We will investigate porous and layered bodies in future work.
Figure \ref{fig:compqstar} shows a range of strength and composition,
from strong rock \citep{Benz99} to weak ice targets. Here, we assume
that changing the equation of state from solid rock to solid ice moves
the $Q_D^*$ curve down in the gravity regime by a factor of the bulk
density, $\sim 3$ between the thick solid line to the thick dashed
line.  There are additional effects from changing equation of state,
e.g., phase changes which modify the shock pressure decay profile;
hence, the offset is not precisely the difference in density \citep[as
shown by][]{Benz99}.  However, at the moment, these effects are within
the numerical error in determining the mass of the largest remnant.
We also focus on 90$^{\circ}$ impact events and note that changing the
impact angle from 90$^\circ$ to and 45$^\circ$ is also within the
range of errors in calculating $Q_D^*$.  Note the small offset between
the 90$^{\circ}$ points (open symbols) and $\sim 45^{\circ}$ thin
lined curves in Fig.~\ref{fig:compqstar}. More oblique impacts ($>
45^{\circ}$) have a more significant effect on $Q_D^*$ \citep{Benz99}.

In the strength regime, we recommend using laboratory experiments on
ice targets \citep[filled circles in Fig.~\ref{fig:compqstar}
from][]{Giblin04, Ryan99, Kato92, Kato95, Arakawa99, Arakawa1995,
  Cintala1985, Kawakami1983, Lange1987, Higa96, Higa98} and assume a
strain rate dependent slope as described in \citet{Housen90}. Note
that the strength regime of \citet{Benz99}'s ice curves are much
stronger than the laboratory data.  The laboratory ice experiments
span a large range in impact velocities (4 to 3500 m~s$^{-1}$) and
initial temperature (77 to 263~K). However, the results are not
monotonic with impact velocity and the scatter likely represents
differences in target preparation.  There is limited data on weak
rocks, but one study on the catastrophic disruption of pyrophyllite
lies between the ice data and the strong rock data in the strength
regime \citep{Takagi84}.

In the gravity regime, the disruption criteria can be estimated for a
range of impact conditions by considering the density of the material,
$\rho$, the impact velocity, and coupling parameter, $\mu$.
\citet{Housen90} derive
\begin{equation}
Q_D^* \sim C \rho R_T^{3\mu} V_i^{-3 \mu +2}, \label{eq:qgrav}
\end{equation}
where $C$ is a constant that is fitted to our simulations and accounts
for the effects of shear strength, $\rho$ in kg~m$^{-3}$, $R_T$ in m,
and $V_i$ in m~s$^{-1}$.  Based on our and previous numerical
experiments, $\mu\sim0.45$.  Note that Eq.~\ref{eq:qgrav} applies only
to hypervelocity collisions where a shock wave develops, which is
expected for most collisions after the end of the accretionary phase
in the solar system. We fit $C=2.6\pm0.9\times 10^{-8}$ for weak
targets (sims.\ 1, 8, 14) and $C=5.2\pm1.7\times 10^{-8}$ for strong
targets (sims.\ 2, 10, 16), using $\rho=2820$~kg~m$^{-3}$ and the
disruption thresholds and critical velocities in
Table~\ref{tab:qstar}.  Note that our weak rock results (thick solid
line in Fig.~\ref{fig:compqstar}) are also applicable to the asteroid
belt.

The intersection of the strength and gravity lines defines the
transition region, which spans a few decades in size.  Using the
functional form of \citet{Benz99},
\begin{equation}
Q_D^* = q_b R^\alpha + \rho q_g R^\beta,
\end{equation}
our recommended parameters for ice targets (with a weak strength) at
1~km~s$^{-1}$ are: $\rho = 930$ kg m$^{-3}$ (low temperature density),
$q_b = 20$ J kg$^{-1}$, $q_g = 3.5\times10^{-6}$ J m$^3$ kg$^{-2}$,
$\alpha = -0.4$, and $\beta = 1.3$ (thick dashed line in
Fig.~\ref{fig:compqstar}).  $q_b$ is centered on the laboratory
experiments, $\rho q_g$ is the value of the gravity section of the
curve at 1 m, $\alpha$ is taken to match \citet{Benz99} which is in
good agreement with \citet{Housenb99}, and $\beta$ is an average
gravity regime exponent based on our and previous work.

In this work, we also developed techniques to study the physical
properties of the largest remnant, which will be used in future work
to compare to observations of Kuiper belt objects. Insights into the
composition and preservation of Kuiper belt objects may be acheived in
future efforts that combine observations and modeling of collisional
processes \citep{Leinhardt08}. Even though the range of shock
pressures is low (e.g., Fig.~\ref{fig:prov}C), because of the expected
physical properties of Kuiper belt objects, some shock modification
may be observed. For example, thermodynamic analyses of shock wave
experiments on 100~K H$_2$O ice indicate that peak shock pressures of
1.6 to 4.1 GPa are required to produce incipient and complete melting,
respectively \citep{Stewart05}. The results from simulation group 19
indicate that catastrophic disruption events in the Kuiper belt reach
conditions for melting ice. The effects of heating from the shock may
produce changes in the composition of chemistry of outer solar system
bodies. Because the reaccumulation process produces a final remnant
with surface materials that have experienced a wide range of peak
shock pressures, we suggest that surfaces of large rubble piles in the
Kuiper belt may have complex, heterogeneous surface features (e.g.,
spectral or albedo variations).

Our analyses indicate that outer solar system bodies may be disrupted
more easily than indicated in previous simulations of catastrophic
disruption. As a result, the collisional evolution of the outer solar
system may have produced more mass loss from collisional grinding
\citep[e.g.,][]{Kenyon04}. At the present time, the relative roles of
dynamical and collisional sculpting of the outer solar system are
poorly understood. The first attempts at combined
dynamical-collisional models have just begun \citep[][and S. Kenyon,
personal communication]{Charnoz07}. \citet{Charnoz07} found it
difficult to reconcile observations of the size distributions of
comets and Kuiper belt objects with their hybrid model of dynamical
and collisional evolution of the outer solar system, and they
suggested that outer solar system bodies could not have lower
disruption energies than found by \citet{Benz99}. In this work, we
find that it is likely that outer solar system bodies require lower
collision energies than assumed by \citet{Charnoz07}, illustrating the
youthful stage of outer solar system collisional studies compared to
studies of the asteroid belt. Recent work on the asteroid belt has
illustrated how combined dynamical-collisional studies may be used to
infer the complex evolution of the inner solar system
\citep{Bottke05a,Bottke05b}. We are making progress toward similarly
sophisticated studies of the outer solar system.

\section{Future Development} \label{sec:future}

The present work focuses on the gravity regime. In the future, we will
further develop the \texttt{CTH}-\texttt{pkdgrav} technique to include
modeling in the strength regime. The ROCK model is able to reproduce
the fracture patterns in laboratory impact cratering and disruption
experiments; however, computational techniques are needed to define
the size distribution of intact fragments. While the largest fragments
may be directly resolved, unresolved fragments will need a statistical
model to determine the size distribution (e.g., the Grady-Kipp model
used in previous work). When the largest intact fragment is greater
than a single \texttt{pkdgrav} particle, methods need to be developed
to bind multiple \texttt{pkdgrav} particles together for the
gravitational reaccumulation portion of the calculation. Modeling the
strength regime will allow study of the transition from the strength
to gravity regime and calculation of full $Q_D^*$ curves for specific
materials.

\section{Conclusions}

We present results from a series of catastrophic
disruption collisions using a new hybrid hydrocode-to-$N$-body
numerical technique. This method has been developed to study the
outcome of collisions between bodies with complex material properties,
such as Kuiper belt objects. Our hybrid method allows us to directly
calculate the mass of the largest gravitationally reaccumulated
remnants. Hence, we also developed techniques to analyze the physical
properties of the reaccumulated remnants from catastrophic collision
events. Understanding the role of collisions in changing the
composition and internal structure of Kuiper belt objects is important
because KBOs are the best representatives of the planetesimals that
accreted into the outer solar system planets.  We find that material
in the largest post-collision remnants experience a large range of
peak shock pressures.  We suggest that large reaccumulated bodies may
have heterogeneous surface features and that interior and exterior
properties may be different.

In this work, we use rocky, asteroid-like targets to facilitate
comparison between our results and that of previous work. We show
that, when similar strength parameters are used, the derived
catastrophic disruption criteria, $Q_D^*$, is in excellent agreement
with previous work \citep{Benz99,Melosh97}. We also demonstrate that
the shear strength of a body has a significant effect on the $Q_D^*$,
even in the gravity regime, and we recommend new catastrophic
disruption criteria for nonporous weak rocky and icy bodies.

\acknowledgements{{\it Acknowledgements.} This work was support by
  NASA grant \#NNG05GH46G. ZML is supported by STFC postdoctoral
  fellowship. We thank D. Richardson and L. Senft for useful
  discussions and feedback. We appreciate the comments from W. Benz
  and an anonymous reviewer that improved this manuscript.}

\bibliography{paper_refs_jun17}

\newpage
\section{Appendix}
Material parameters used with the ROCK model in \texttt{CTH} are
presented in Table~\ref{tab:rockparams}.  The ROCK model is based upon
the strength model developed by \citet{Collins04} for the
\texttt{SALEB} hydrocode and implemented into \texttt{CTH} by
\citet{Senft07}. Note that the total damage, $D$, is the sum of the
shear and tensile damage and limited to a maximum value of 1.

\newpage
\begin{table}
\caption{Principle parameters for different strength models.
  Simulations denoted by R use the ROCK model (otherwise GEO
  model). Note that $Y_0=10^7$ for strong ROCK models; full parameters
  for the ROCK model are given in the Appendix. 
\label{tab:strength}}
\begin{tabular}{lllll}
\tableline
Strength & $Y_0$ & $Y_M$ & $Y_T$ & Sim. \\
Model & Pa & Pa & Pa & Num.\\
\tableline
Hydro & 0 & 0 & $10^5$ & 9, 15 \\
Weak & $10^6$ & $10^7$ & $10^5$ & 1, 4R, 8, 14, 17, 18, 19 \\
Medium & $10^7$ & $10^8$ & $10^6$ & 13 \\
Strong & $10^6$ & $3.5\times10^9$ & $10^5$ & 2, 6R, 7R, 10, 11R,
12R, 16 \\
StrongHT & $10^6$ & $3.5\times10^9$ & $10^7$ & 3, 5R \\
\tableline
\end{tabular}
\end{table}

\newpage

\tablewidth{7in}
\tabletypesize{\footnotesize}
\begin{deluxetable}{rlccccccccrrl}
\rotate
\tablecaption{Summary of simulation parameters and results. \label{tab:data}}
\tablecolumns{13}
\tabletypesize{\footnotesize}
\tablehead{\colhead{Sim.} & \colhead{Sim.$^{\dagger}$} & \colhead{Mat.$^{\dagger}$} & \colhead{\texttt{CTH}} & \colhead{$R_T$} & \colhead{$R_P$} & \colhead{$\alpha^{\ddagger}$} & \colhead{$V_{i}$}& \colhead{Strength} & \colhead{$t_{ho}$} & \colhead{$N_{init}^{\ddagger}$} & \colhead{$N_{lr}^{\ddagger}$} & \colhead{$M_{lr}/M_{T}^{\ddagger}$} \\
\colhead{Num.} & \colhead{Type} & \colhead{} & \colhead{Dim.} & \colhead{[km]} & \colhead{[km]} & \colhead{[$^\circ$]} & \colhead{[km s$^{-1}$]} & \colhead{Model} & \colhead{[s]} & \colhead{} & \colhead{} 
}

\startdata

1a & G\_merge & B & 2 & 2 & 0.17 & 90 & 1.3 & weak & 2.5 & 3727 & 24 & 0.08\tablenotemark{*} \\
1b & G\_merge & B & 2 & 2 & 0.17 & 90 & 1.0 & weak & 2.5 & 3728 & 278 & 0.31 \\
1c & G\_merge & B & 2 & 2 & 0.17 & 90 & 0.7 & weak & 2.5 & 2166 & 349 & 0.62 \\ 
2a & G\_merge & B & 2 & 2 & 0.17 & 90 & 2.0 & strong & 2.5 & 4829 & 239 & 0.06\tablenotemark{*} \\
2b & G\_merge & B & 2 & 2 & 0.17 & 90 & 1.5 & strong & 2.5 & 3257 & 645 & 0.39 \\
2c & G\_merge & B & 2 & 2 & 0.17 & 90 & 1.0 & strong & 2.5 & 4026 & 2160 & 0.81 \\ 
3a & G\_merge & B & 2 & 2 & 0.17 & 90 & 3.0 & strongHT & 2.5 & 16311 & 1852 & 0.37 \\
3b & G\_merge & B & 2 & 2 & 0.17 & 90 & 2.5 & strongHT & 2.5 & 13283 & 2023 & 0.50 \\
3c & G\_merge & B & 2 & 2 & 0.17 & 90 & 2.0 & strongHT & 2.5 & 11551 & 2908 & 0.62 \\
4a & R\_merge & B & 2 & 2 & 0.17 & 90 & 1.3 & weak & 2.5 & 22256 & 624 & 0.06\tablenotemark{*} \\
4b & R\_merge & B & 2 & 2 & 0.17 & 90 & 1.0 & weak & 2.5 & 20861 & 3978 & 0.44 \\
4c & R\_merge & B & 2 & 2 & 0.17 & 90 & 0.7 & weak & 2.5 & 19417 & 9580 & 0.79 \\
5a & R\_merge & B & 2 & 2 & 0.17 & 90 & 2.0 & strong & 2.5 & 5959 & 216 & 0.17 \\
5b & R\_merge & B & 2 & 2 & 0.17 & 90 & 1.5 & strong & 2.5 & 3375 & 989 & 0.53 \\
5c & R\_merge & B & 2 & 2 & 0.17 & 90 & 1.0 & strong & 2.5 & 2863 & 1379 & 0.79 \\
6a & R\_merge & B & 2 & 2 & 0.17 & 90 & 3.0 & strongHT & 2.5 & 11194 & 1159 & 0.32 \\
6b & R\_merge & B & 2 & 2 & 0.17 & 90 & 2.5 & strongHT & 2.5 & 5092 & 266 & 0.21 \\
6c & R\_merge & B & 2 & 2 & 0.17 & 90 & 2.0 & strongHT & 2.5 & 5132 & 1220 & 0.58 \\
7a & R\_merge & B & 2 & 2 & 0.15 & 90 & 3.0 & strongHT & 2.5 & 4820 & 747 & 0.40 \\
7b & R\_merge & B & 2 & 2 & 0.12 & 90 & 3.0 & strongHT & 2.5 & 3092 & 619 & 0.68 \\
7c & R\_merge & B & 2 & 2 & 0.10 & 90 & 3.0 & strongHT & 2.5 & 1882 & 483 & 0.88 \\
8a & G\_merge & B & 2 & 10 & 0.83 & 90 & 3.7 & weak & 12 & 25940 & 140 & 0.05\tablenotemark{*} \\ 
8b & G\_merge & B & 2 & 10 & 0.83 & 90 & 2.8 & weak & 12 & 14538 & 228 & 0.29 \\
8c & G\_merge & B & 2 & 10 & 0.83 & 90 & 1.9 & weak & 12 & 8020 & 1350 & 0.76 \\
9a & G\_merge & B & 2 & 10 & 0.83 & 90 & 1.9 & hydro & 20 & 5491 & 1119 & 0.40 \\
9b & G\_merge & B & 2 & 10 & 0.83 & 90 & 1.0 & hydro & 20 & 7141 & 1112 & 0.81 \\
10a & G\_merge & B & 2 & 10 & 0.83 & 90 & 5.6 & strong & 12 & 33500 & 3668 & 0.36 \\
10b & G\_merge & B & 2 & 10 & 0.83 & 90 & 4.6 & strong & 12 & 6216 & 371 & 0.53 \\ 
10c & G\_merge & B & 2 & 10 & 0.83 & 90 & 3.7 & strong & 12 & 9614 & 3328 & 0.66 \\
11a & R\_merge & B & 2 & 10 & 0.83 & 90 & 5.6 & strongHT & 25 & 23015 & 1059 & 0.35 \\
11b & R\_merge & B & 2 & 10 & 0.83 & 90 & 4.6 & strongHT & 15 & 14929 & 1642 & 0.53 \\
11c & R\_merge & B & 2 & 10 & 0.83 & 90 & 3.7 & strongHT & 12 & 12993 & 2003 & 0.72 \\
12a & R\_merge & B & 2 & 10 & 1.5 & 90 & 3.0 & strongHT & 12 & 16616 & 534 & 0.04\tablenotemark{*} \\
12b & R\_merge & B & 2 & 10 & 1.2 & 90 & 3.0 & strongHT & 12 & 11773 & 2254 & 0.40 \\
12c & R\_merge & B & 2 & 10 & 1.0 & 90 & 3.0 & strongHT & 12 & 6198 & 4752 & 0.64 \\
13a & G\_merge & B & 2 & 10 & 0.83 & 90 & 3.7 & med & 12 & 29601 & 406 & 0.13 \\ 
13b & G\_merge & B & 2 & 10 & 0.83 & 90 & 2.8 & med & 12 & 7445 & 971 & 0.59 \\ 
13c & G\_merge & B & 2 & 10 & 0.83 & 90 & 1.9 & med & 12 & 6693 & 1938 & 0.90 \\
14a & G\_merge & B & 2 & 50 & 14 & 90 & 1.8 & weak & 60 & 9866 & 161 & 0.04\tablenotemark{*} \\ 
14b & G\_merge & B & 2 & 50 & 14 & 90 & 1.4 & weak & 60 & 11998 & 2950 & 0.40 \\ 
14c & G\_merge & B & 2 & 50 & 14 & 90 &1.2 & weak & 60 & 8883 & 1775 & 0.60 \\
14d & G\_merge & B & 2 & 50 & 14 & 90 & 1.0 & weak & 60 & 6761 & 1943 & 0.73 \\ 
15a & G\_merge & B & 2 & 50 & 14 & 90 & 1.4 & hydro & 60 & 14244 & 1118 & 0.11 \\ 
15b & G\_merge & B & 2 & 50 & 14 & 90 & 1.0 & hydro & 60 & 11425 & 5632 & 0.74 \\ 
16a & G\_merge & B & 2 & 50 & 14 & 90 & 3.0 & strong & 60 & 11753 & 171 & 0.02\tablenotemark{*} \\ 
16b & G\_merge & B & 2 & 50 & 14 & 90 & 2.0 & strong & 60 & 11565 & 1949 & 0.34 \\ 
16c & G\_merge & B & 2 & 50 & 14 & 90 & 1.5 & strong & 60 & 9803 & 4873 & 0.74 \\ 
17a & G\_merge & B & 3 & 50 & 14 & 90 & 1.2 & weak & 60 & 7469 & 1513 & 0.57 \\ 
17b & G\_bounce & B & 3 & 50 & 14 & 90 & 1.2 & weak & 60 & 7469 & 93 & 0.21\tablenotemark{*} \\ 
17c & G\_loweps & B & 3 & 50 & 14 & 90 & 1.2 & weak & 60 & 7469 & 63 & 0.22\tablenotemark{*} \\ 
18a & G\_merge & B & 3 & 50 & 14 & 45 & 1.8 & weak & 60 & 6877 & 1829 & 0.74 \\ 
18b & G\_bounce & B & 3 & 50 & 14 & 45 & 1.8 & weak & 60 & 6877 & 278 & 0.45 \\ 
18c & G\_loweps & B & 3 & 50 & 14 & 45 & 1.8 & weak & 60 & 6877 & 309 & 0.46 \\ 
19a & G\_merge & I & 3 & 50 & 8.5 & 45 & 3.0 & weak & 60 & 2954 & 1140 & 0.49 \\
19b & G\_merge & I & 3 & 50 & 8.5 & 45 & 3.0 & weak & 60 & 2854 & 1092 & 0.49 \\
19c & G\_merge & I & 3 & 50 & 8.5 & 45 & 3.0 & weak & 60 & 4822 & 1160 & 0.49 \\
19d & G\_merge & I & 3 & 50 & 8.5 & 45 & 3.0 & weak & 60 & 5035 & 1197 & 0.49 \\
19e & G\_bounce & I & 3 & 50 & 8.5 & 45 & 3.0 & weak & 60 & 4822 & 377 & 0.24  
\enddata 

\tablenotetext{*}{Under-resolved largest remnant.} 
\tablenotetext{\dagger}{G -- GEO model; R -- ROCK model; B -- basalt; I -- ice.}
\tablenotetext{\ddagger}{$\alpha$ -- impact angle measured from the
  tangent plane; $N_{init}$ -- number of \texttt{pkdgrav} particles at
handoff; $N_{lr}$ -- number of \texttt{pkdgrav} particles in the
largest remnant; $M_{lr}/M_{T}$ -- mass of largest post-collision
remnant / target mass.}

\end{deluxetable}

\newpage

\begin{table}
\caption{Catastrophic Disruption Threshold\label{tab:qstar}}
\begin{tabular}{lll}
\tableline
Sim.\ & $V_{\rm crit}$ [km/s] & $Q_D^*$ [J/kg]\\
\tableline
1 & $0.83\pm .1$ & $2.1(\pm 0.5)\times 10^2$\\
2 & $1.4\pm .1$ & $6.3 (\pm 0.8)\times 10^2$\\
3 & $2.5\pm .3$ & $1.9 (\pm 0.4)\times 10^3$\\
4 & $0.97\pm .05$ & $2.9 (\pm 0.3)\times 10^2$\\
5 & $1.5\pm .1$ & $7.4 (\pm 0.9)\times 10^2$\\
6 & $1.9\pm .5$ & $1.1 (\pm 0.6)\times 10^3$\\
8 & $2.50\pm .04$ & $1.79 (\pm 0.06)\times 10^3$\\
9\tablenotemark{\dagger} & 1.7 & $8.5 \times 10^2$\\
10 & $4.8\pm .3$ & $6.6 (\pm 0.9) \times 10^3$\\
11 & $4.9\pm .3$ & $6.8 (\pm 0.8) \times 10^3$\\
13 & $3.0\pm .1$ & $2.5 (\pm 0.2) \times 10^3$\\
14 & $1.30\pm .07$ & $1.8 (\pm 0.2) \times 10^4$\\
15\tablenotemark{\dagger} & 1.2 & $1.5 \times 10^4$\\
16 & $1.9\pm .2$ & $4.0 (\pm 0.7) \times 10^4$\\
\tableline
\end{tabular}
\tablenotetext{\dagger}{No error bars on simulation sets that contain only two simulations. All other simulation sets have three simulations.}
\end{table}

\begin{table}
\caption{Catastrophic Disruption Threshold\label{tab:qstarcv}}
\begin{tabular}{lll}
\tableline
Sim.\ & $R_{\rm crit}$ [km] & $Q_D^*$ [J/kg]\\
\tableline
7 & $0.141\pm .003$ & $1.6 (\pm 0.2) \times 10^2$\\
12 & $1.13\pm .06$ & $7 (\pm 1) \times 10^3$\\
\tableline
\end{tabular}
\end{table}


\newpage
\renewcommand{\thetable}{A\arabic{table}}
\setcounter{table}{0}
\begin{table}
\caption{ROCK model parameters. \label{tab:rockparams}}
\begin{tabular}{llllll}
\tableline
Parameter &   Weak & Strong   &   StrongHT  & Description \\
\tableline
$Y_0$ (MPa) &  1   & 10      &   10         & shear strength at zero pressure \\
$Y_M$ (GPa) &  0.01 & 3.5     &   3.5       & shear strength Von Mises limit \\
$\phi_i$    &  1.2 & 1.2      &   1.2        & coefficient of internal friction (D=0) \\
$Y_c$ (Pa) &  0  & 0        &   0           & cohesion (D=1) \\
$\phi_d$    &  0.6 & 0.6      &   0.6        & coefficient of friction (D=1)\\
$Y_T$ (MPa) & -0.1 & -0.1     &  -10        & tensile strength \\
$T_m$ (K)  &  1392 & 1392     &   1392      & melting temperature at ambient pressure \\
$\xi$      &  1.2 & 1.2      &   1.2        & thermal softening parameter \\
$P_{bd}$ (GPa) & 0.003 & 2.94 &   2.94      & brittle-ductile transition pressure  \\
$P_{bp}$ (GPa) & 0.004 & 4.11 &   4.11      & brittle-plastic transition pressure \\
$\nu$           & 0.23 & 0.23 & 0.23 & Poisson's ratio \\
$D_{S0}$        & 0 &   0    &   0          & initial shear damage \\
$D_{T0}$        & 0 &   0    &   0          & initial tensile damage \\
\tableline
\end{tabular}
\end{table}

\newpage

%
\begin{figure}
\centerline{\includegraphics[scale=0.8]{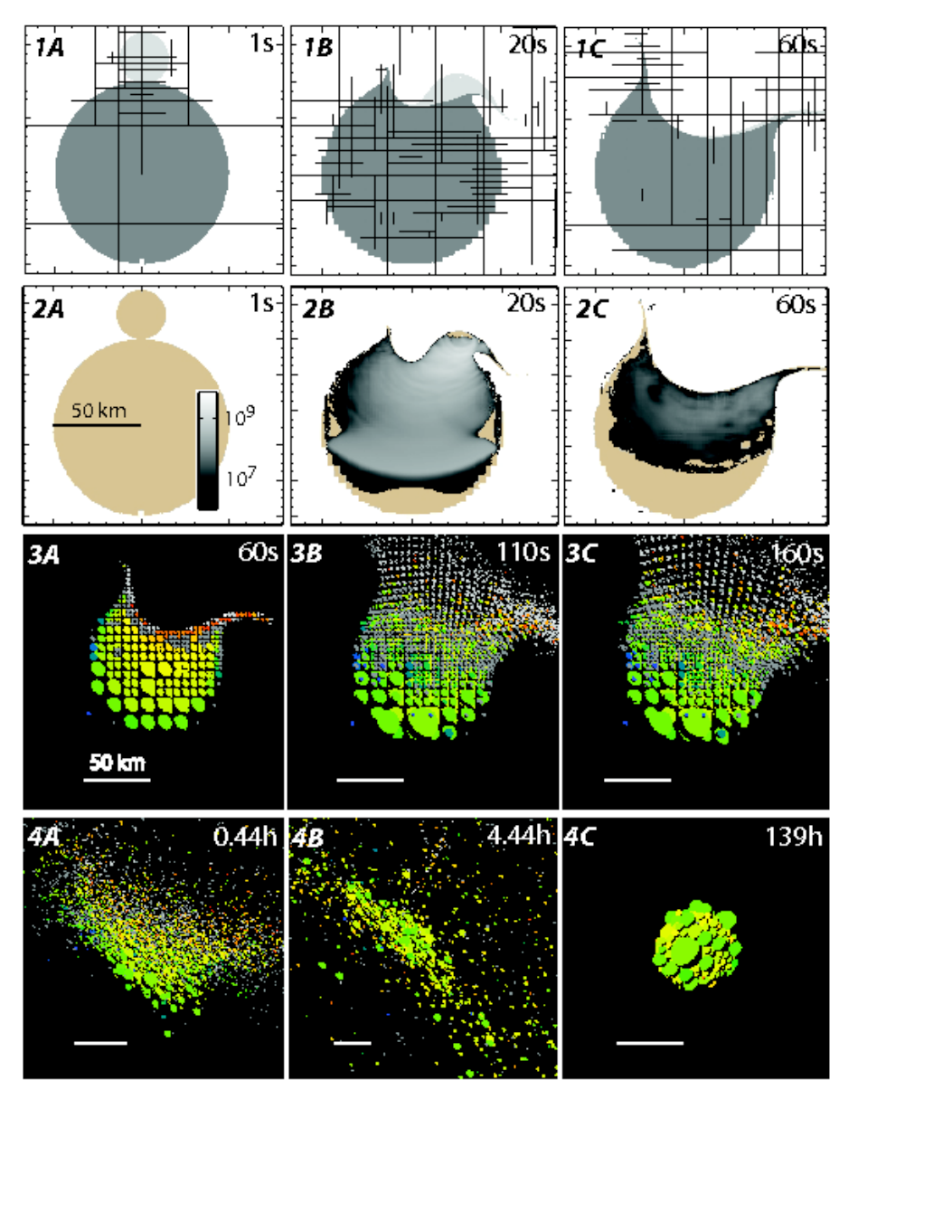}} 
\caption{}
\end{figure}

\addtocounter{figure}{-1}
\begin{figure}
\caption{{\small Example of a 3D, 45 degree, impact
    between two basalt spheres ($R_{P}=14$ km, $R_{T}=50$ km, $v_{i}$
    = 1.8 km s$^{-1}$) using the \texttt{CTH-pkdgrav} technique. Rows
    1-2: \texttt{CTH} calculation (0-60 s) in cross-section along the
    $y = 0$ plane. Rows 3-4: \texttt{pkdgrav} calculation (60 s - 140
    h). Row 1 depicts the projectile (light grey), the target (dark
    grey) and the adaptive mesh.  Row 2 shows the projectile and
    target (beige) and the pressure due to the impact (grey scale in
    Pa). 3A is a cross section ($y = 0$ plane) of the target at 60 s
    after handoff from \texttt{CTH} to \texttt{pkdgrav}. Colors
    represent the peak pressure attained during the impact
    (logarithmic range of 0.01 to 7 GPa).  3B-4C show the entire 3D
    object, zooming out as the disruption process proceeds. Scale bar
    is 50 km.  The last frame shows only the largest reaccumulated,
    post-collision remnant which equilibrates to $45\%$ of the
    target mass in this simulation (sim 18b).
 \label{fig:strip}}}
\end{figure}

\begin{figure}
\noindent \includegraphics[scale=0.6]{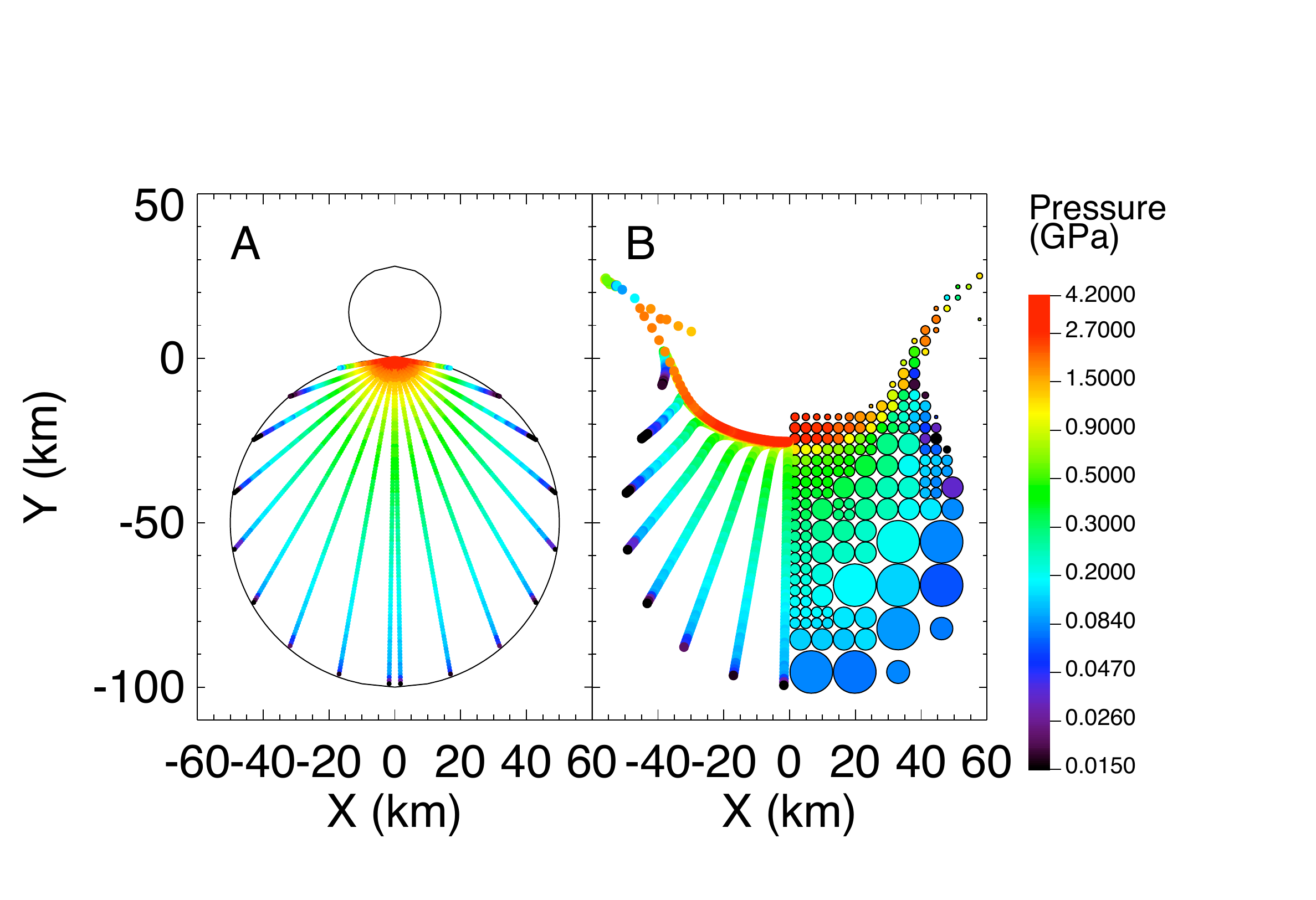}
\caption{
  {\small A. Initial tracer locations. B. Tracer locations and
    \texttt{pkdgrav} particles at handoff. Tracer and particles are
    color coded by peak pressure attained during the impact event.
    Note that projectile material lines the crater floor, shown by
    \texttt{pkdgrav} particles but not tracers. \label{fig:2dtrace}}}
\end{figure}

\begin{figure}
\includegraphics[scale=0.6]{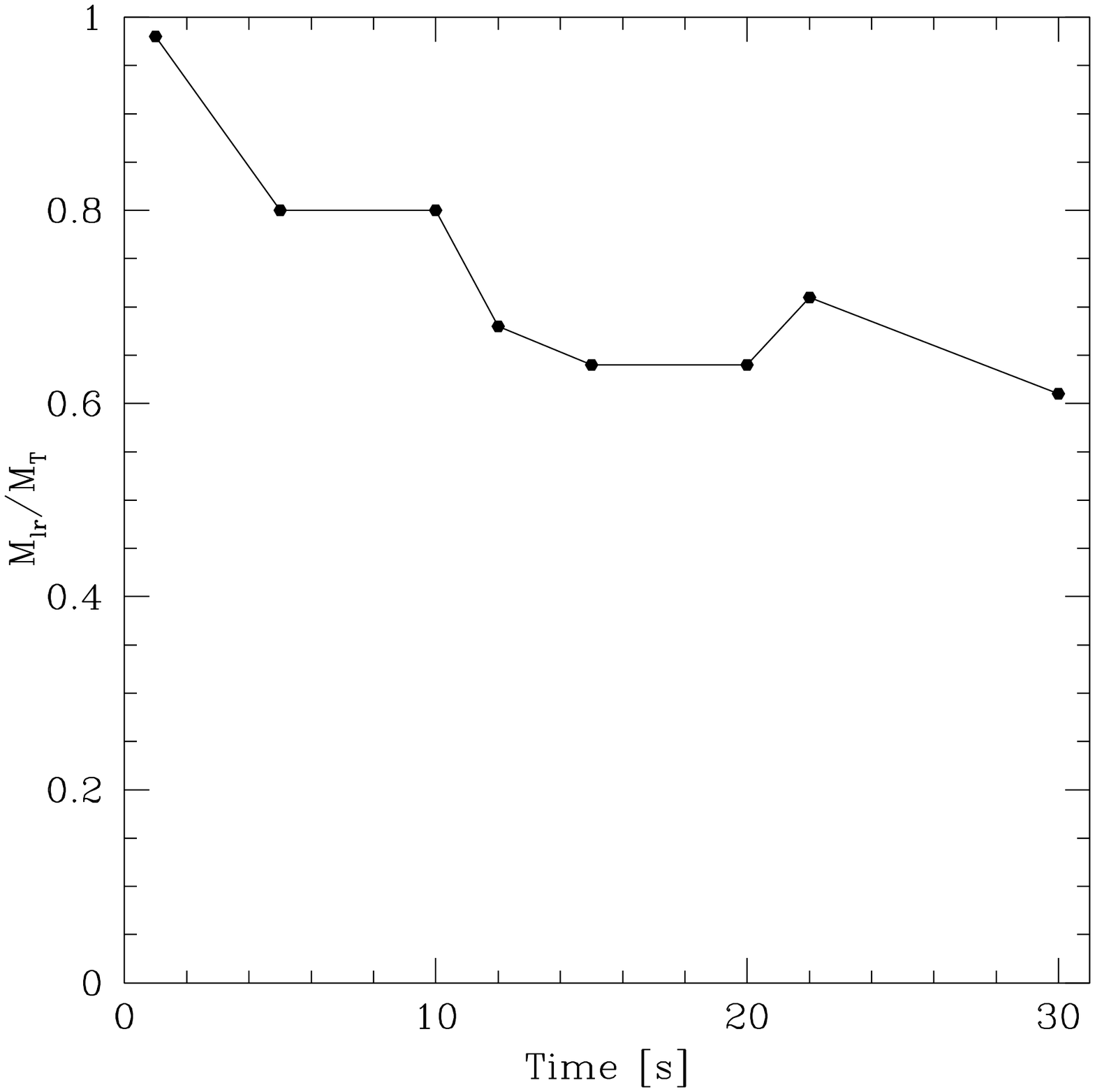} 
\caption{
  {\small Variation in mass of the largest post-collision remnant with
    handoff time from \texttt{CTH} to \texttt{pkdgrav}. $R_T=10$~km,
    $R_P=0.83$~km, and $V_i=1.9$~km s$^{-1}$ (sim.~8c).
    \label{fig:handoff}}}
\end{figure}

\begin{figure}
\includegraphics[scale=0.4]{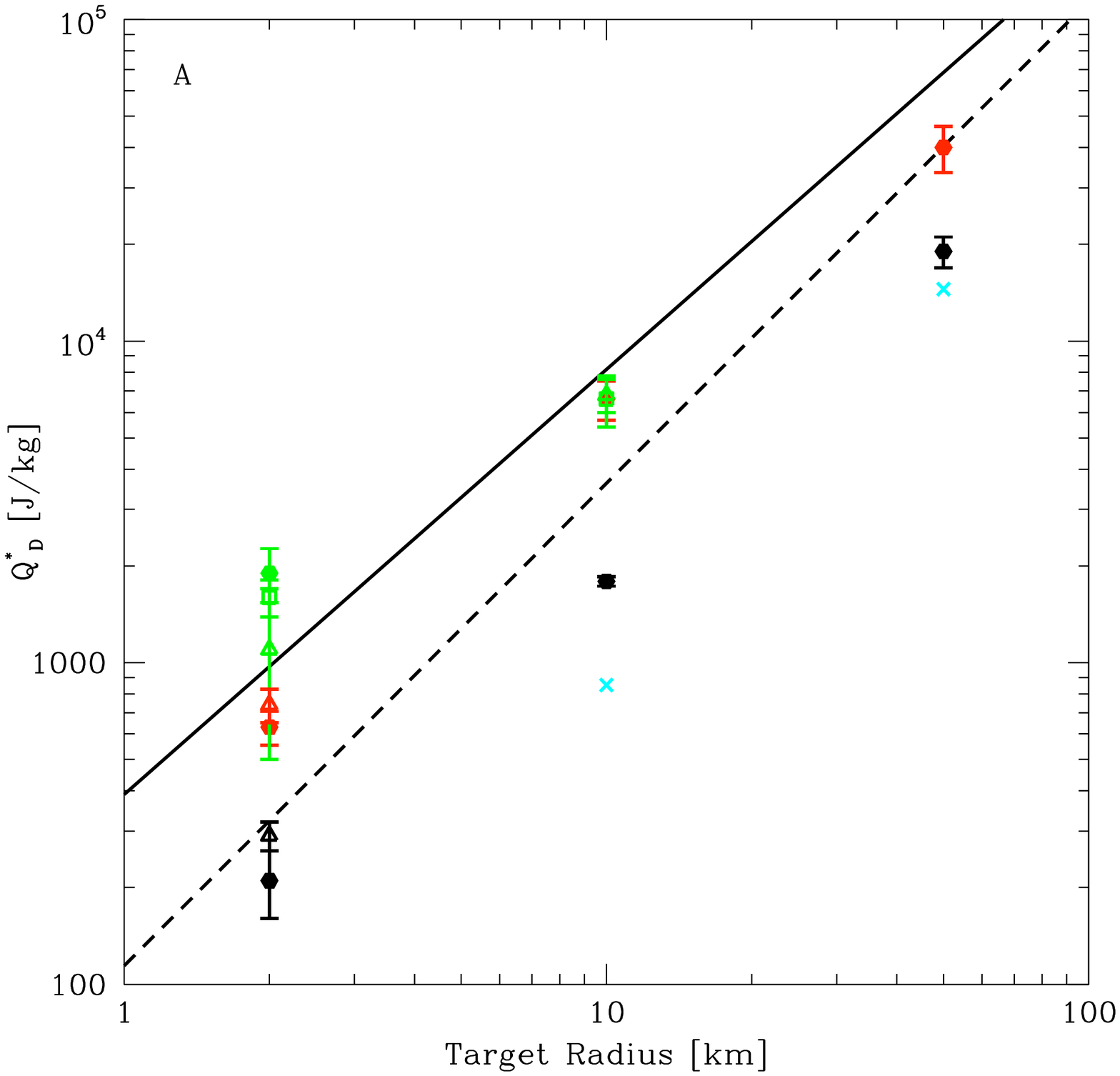} \includegraphics[scale=0.4]{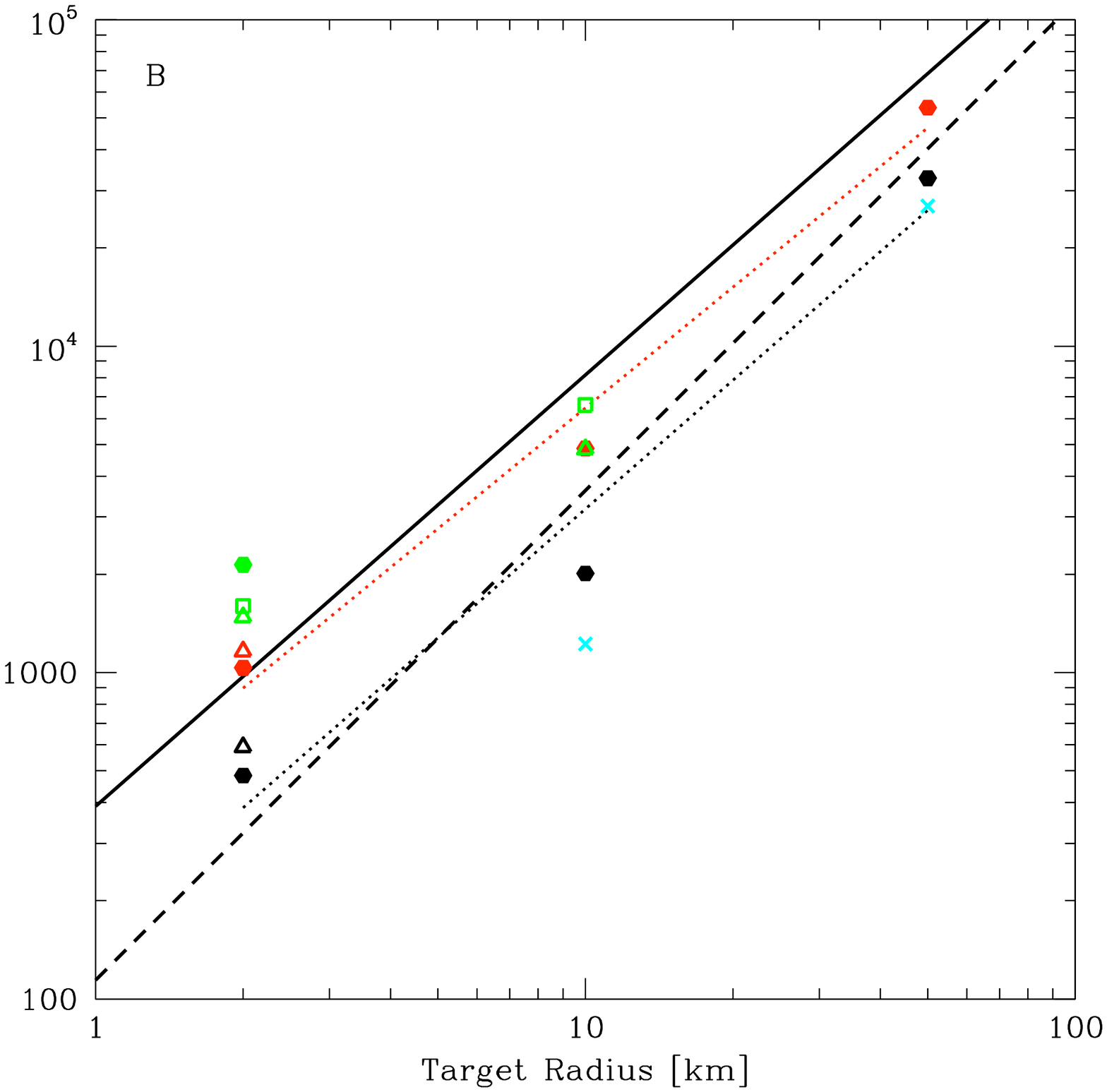} 
\caption{
  {\small A.\ Catastrophic disruption criteria ($Q^*_D$) for basalt
    targets with varying strength models and impact velocities
    (Tables~\ref{tab:qstar} \& \ref{tab:qstarcv}).  B.\ $Q_D^*$ data
    adjusted to $V_i = 3$~km s$^{-1}$.  Solid lines: fit to
    3~km~s$^{-1}$, 90$^{\circ}$ impacts \citep[slope = 1.26,][]{Benz99}.
    Dashed lines: \citet{Melosh97} Eq.\ 5 with $V_i=3$~km s$^{-1}$
    (slope = 1.5).  Data points, with $1\sigma$ errors, from this
    work: black dots (Sims.\ 1, 8, 14), triangle (4R) -- weak
    targets (black dotted line, slope = 1.2); blue $\times$ (9, 15) --
    hydrodynamic targets; red dots (2, 10, 16) and open triangle (5R)
    -- strong targets (red dotted line, slope = 1.3); green point (3),
    open triangles (6R, 11R), open squares (7R, 12R) -- strongHT
    targets.
    \label{fig:qstar}}}
\end{figure}

\begin{figure}
\includegraphics[scale=0.6]{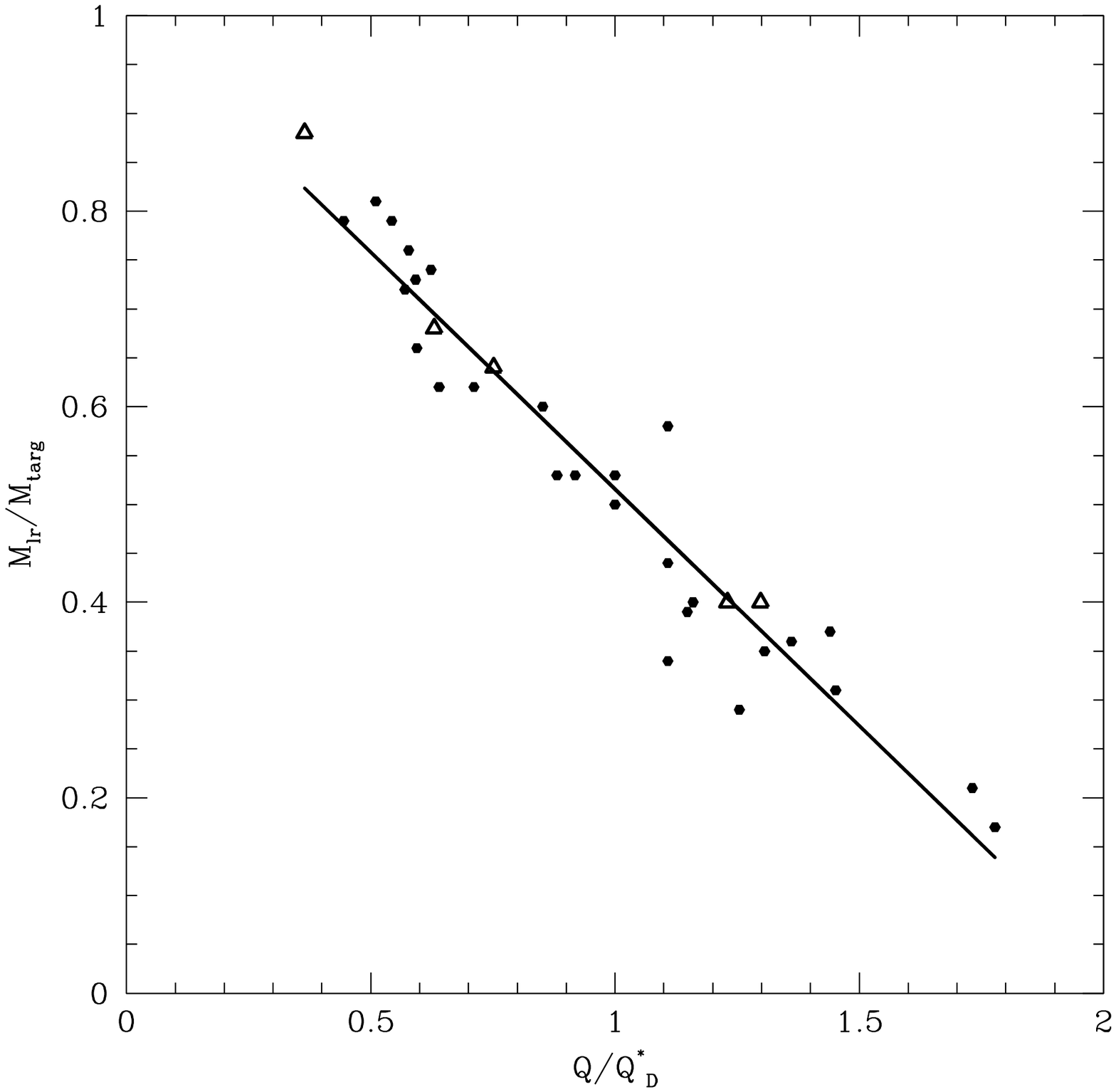} 
\vspace{-3cm}
\caption{
  {\small Mass of the largest post-collision remnant normalized to the
    mass of the target versus the specific impact energy normalized to
    the catastrophic disruption threshold. Each 2D simulation
    with constant mass ratio from Table \ref{tab:data} is represented
    by solid hexagon. Each simulation with varying mass ratio is
    represented by an open triangle. A least squares fit to the data
    produces a slope of $-0.48 \pm 0.02$ (black line). The same linear
    relationship holds for strong to hydrodynamic basalt targets and
    impact velocities from 0.7 to 5.6~km~s$^{-1}$.
\label{fig:mlrvsQ}}}
\end{figure}

\begin{figure}
\includegraphics[scale=0.7]{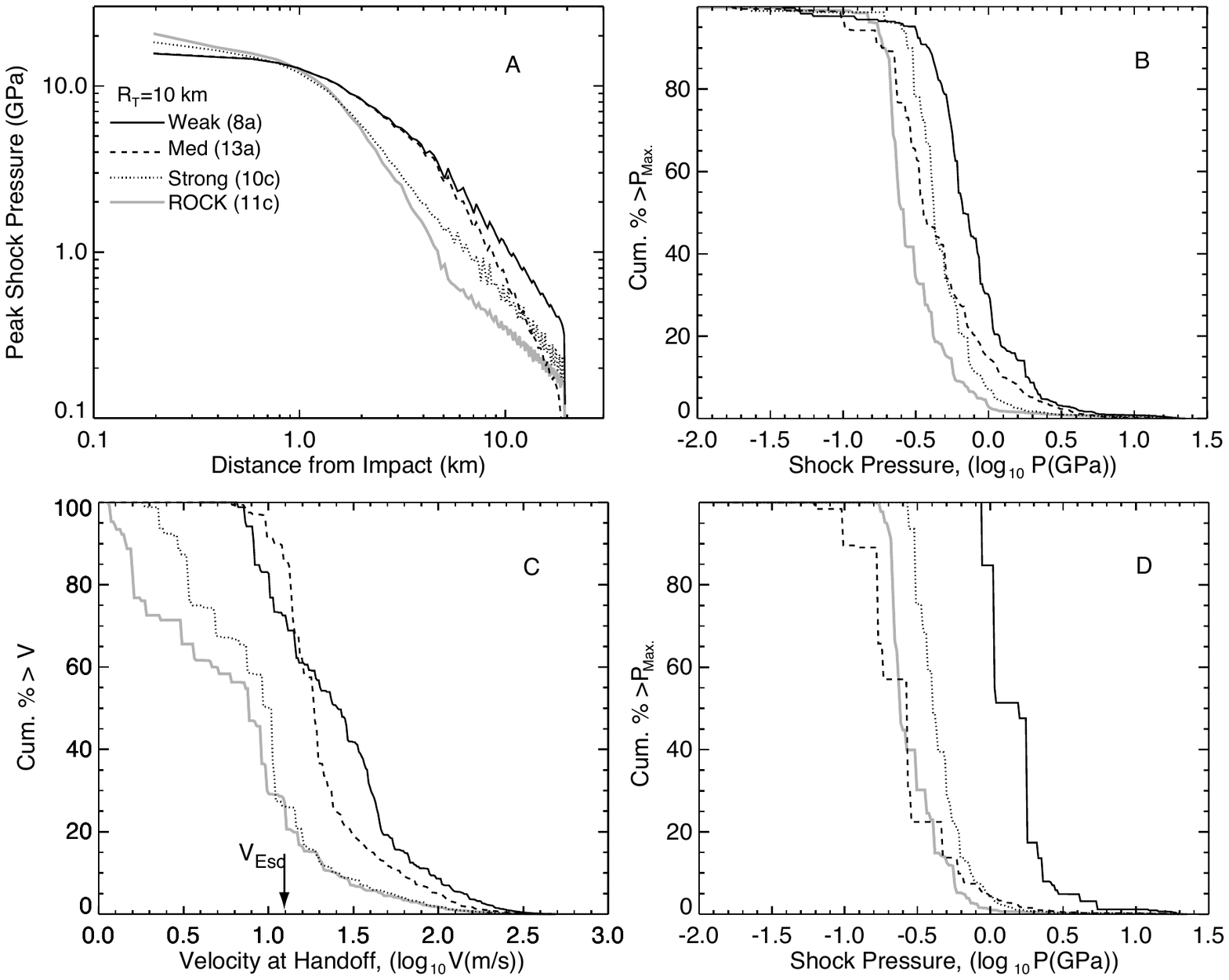}
\vspace{-5cm}
\caption{
  {\small A.\ Peak shock pressure as a function of
    distance (adjacent to the centerline) in the target for
    simulations with varying shear strengths but identical impact
    conditions. In all cases, an 0.83-km
    radius projectile hit a 10-km radius target at 3.7 km~s$^{-1}$.
    B.\ Cumulative plot of the maximum shock pressure distribution (in
    mass fraction) in the total mass. C.\ Cumulative plot of the
    velocity distribution at handoff to \texttt{pkdgrav}.  D.\ 
    Cumulative maximum shock pressure distribution of material
    reaccumulated into the largest remnant.
\label{fig:strength}}}
\end{figure}

\begin{figure}
\includegraphics[scale=0.6]{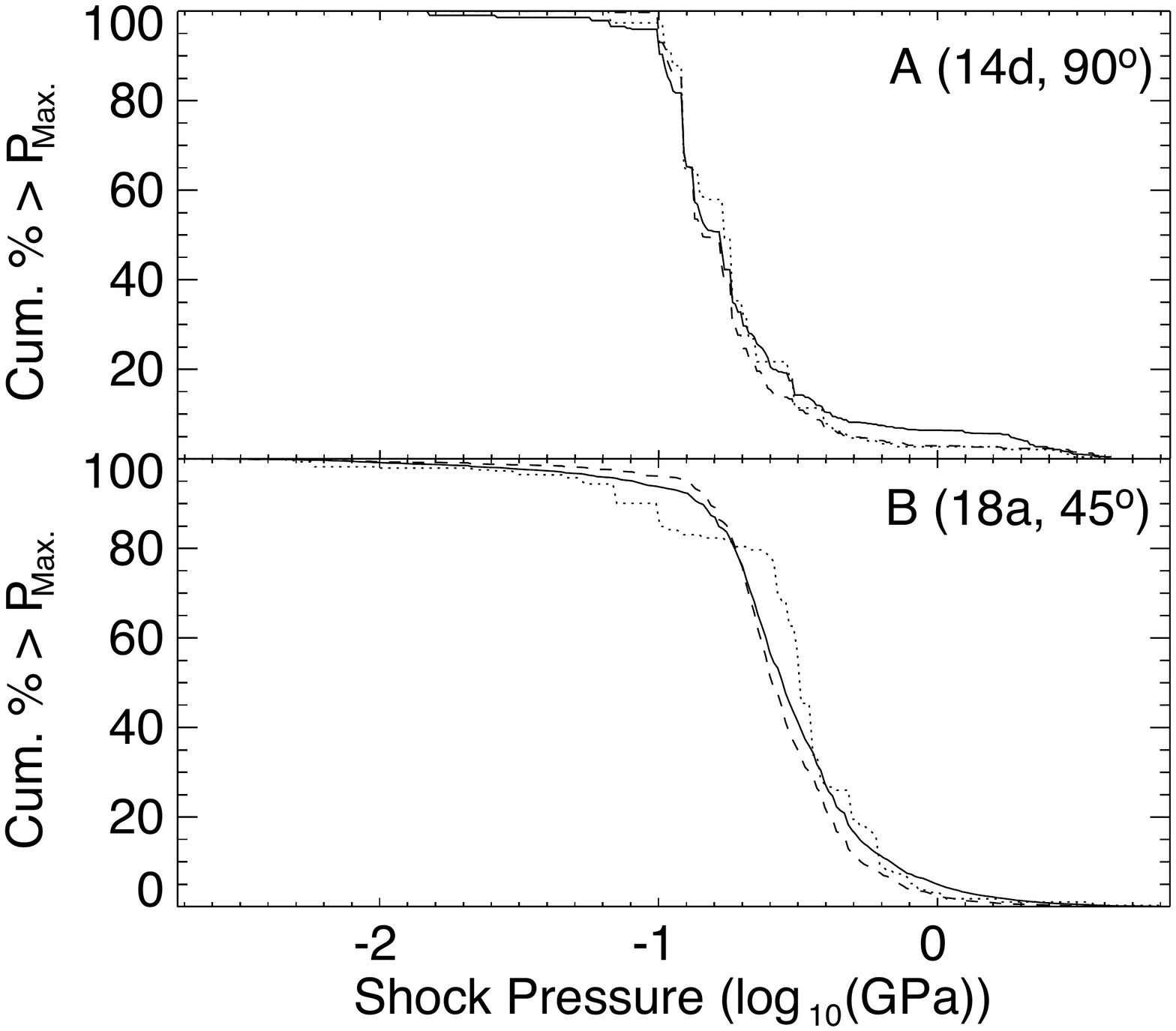}
\caption{
 {\small A. A 2D
    90$^{\circ}$ impact into a 50 km target at $V_i = 1$~km s$^{-1}$,
    $M_{lr}/M_T = 0.73$ (sim.\ 14d). B. A 3D 45$^{\circ}$ impact
    into the same target at $V_i = 1.8$~km~s$^{-1}$, $M_{lr}/M_T
    = 0.74$ (sim.\ 18a). Solid line is total mass, dashed line is the
    largest remnant, dotted line is second largest remnant.
\label{fig:angeff}}}
\end{figure}

\begin{figure}
  \includegraphics[scale=.8]{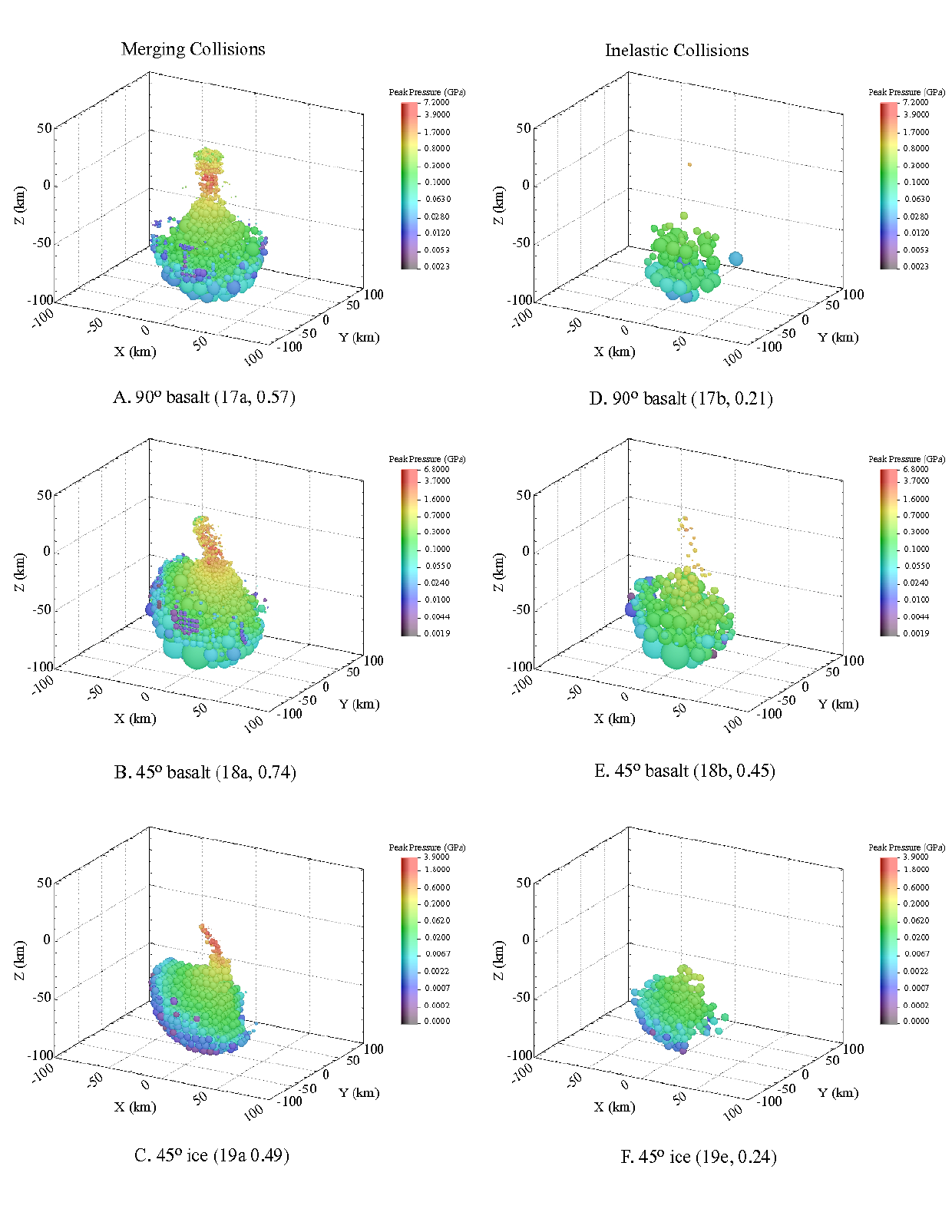}
\caption{}
\end{figure}

\addtocounter{figure}{-1}
\begin{figure}
\caption{
  {\small Provenance plots for the largest remnant in 3D simulations
    of collisions into 50-km radius targets.  Each sphere corresponds
    to a \texttt{pkdgrav} particle; colors represent peak shock
    pressure. Left column: perfect merging results. Right column:
    inelastic collision (bouncing) results. Simulation numbers and
    $M_{lr}/M_T$ are noted in parenthesis. \label{fig:prov}}}
\end{figure}

\begin{figure}
 \includegraphics[scale=0.85]{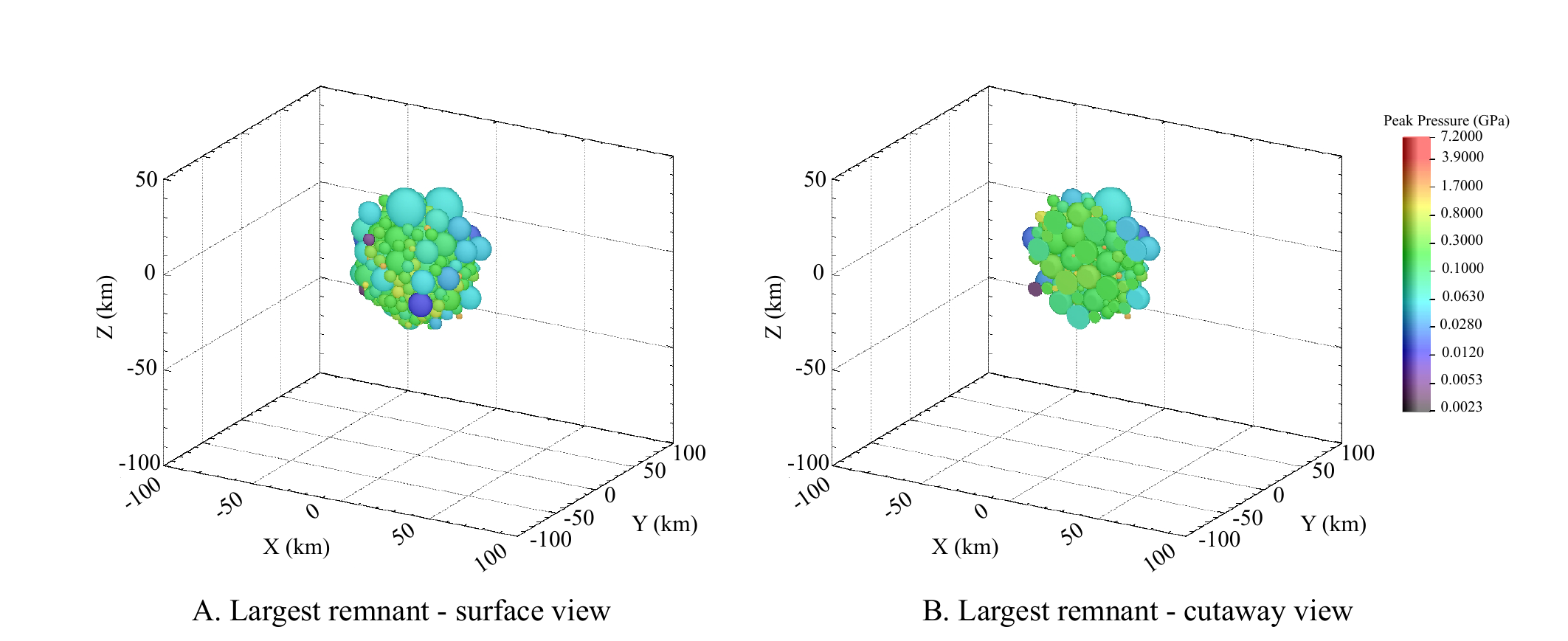}
\caption{
  {\small The largest remnant, $M_{lr}/M_T=0.21$, from 3D simulation of
    45$^{\circ}$ impact onto 50-km basalt target (sim.~17b). A. View
    of surface. B. View of interior, showing rear hemisphere cut
    along the $Y=0$ plane. Colors correspond to peak shock pressure.
\label{fig:surf}}}
\end{figure}

\begin{figure}
 \includegraphics[scale=0.75]{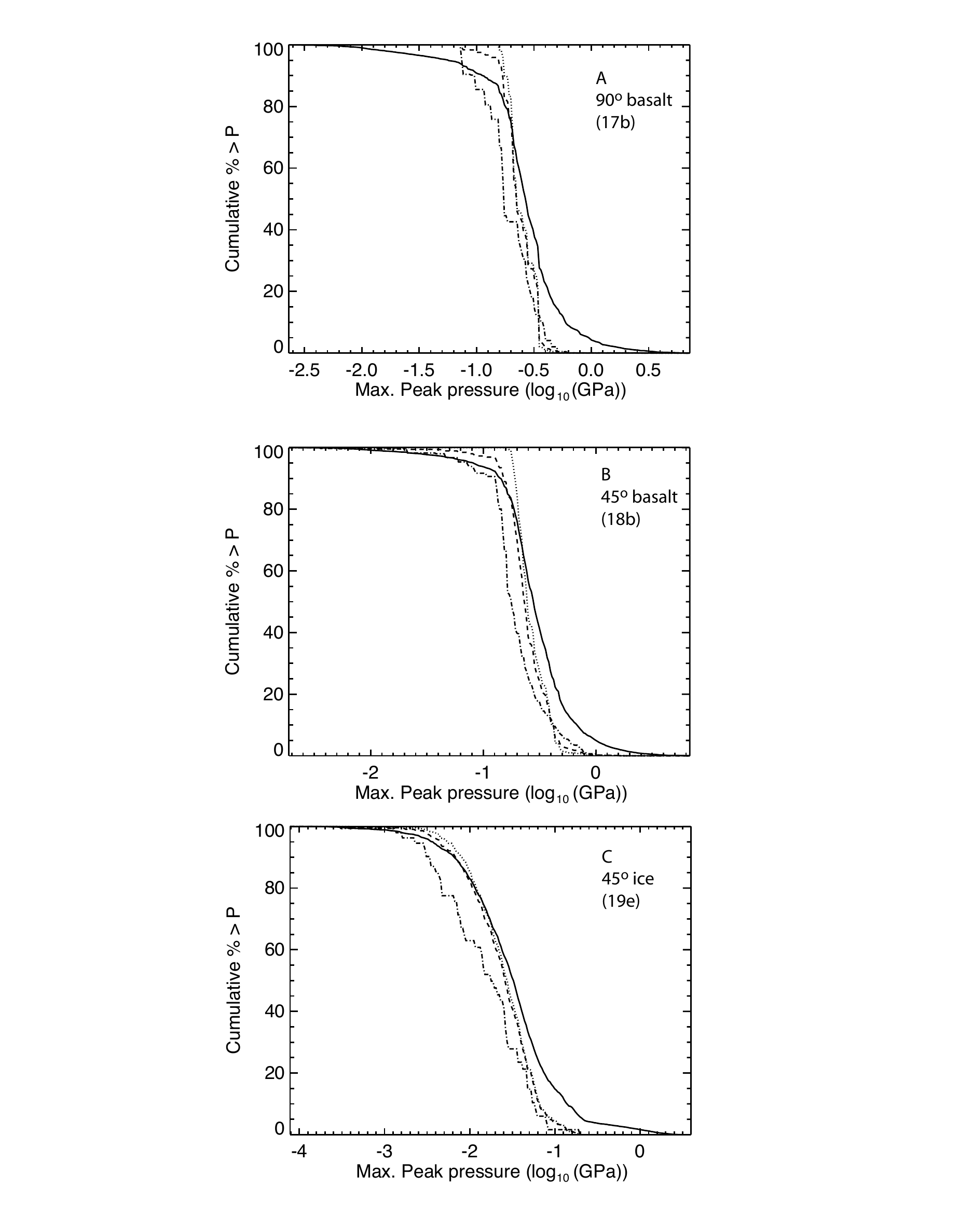}
\caption{
  {\small Interior versus exterior shock pressure history of the
    largest remnant from 3D inelastic bouncing simulations of 50-km
    radii targets. Solid line: all material; dashed line: all of
    largest remnant; dash-dot: exterior of largest remnant ($> 0.75
    R_{lr}$); dotted: interior of largest remnant ($< 0.75 R_{lr}$).
\label{fig:intvsext}}}
\end{figure}

\begin{figure}
\includegraphics[scale=1.]{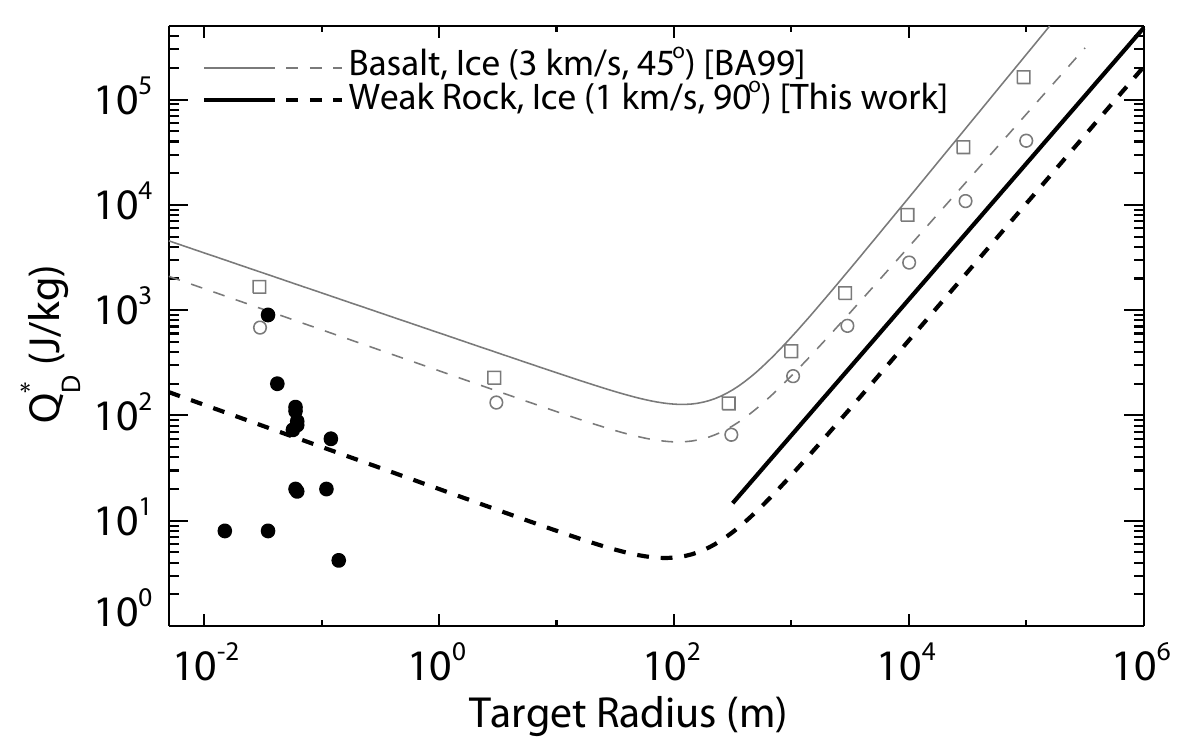} 
\caption{
  {\small Suggested range of $Q_D^*$ curves for small outer solar
    system bodies, such as Kuiper belt objects. Solid lines and
    squares -- rock; dashed lines and circles -- ice.  \citet{Benz99}
    results: thin grey lines -- 3~km~s$^{-1}$ angle averaged fits and
    open symbols -- 90$^{\circ}$ simulations.  Closed circles --
    90$^{\circ}$ laboratory catastrophic disruption data on ice (see
    text).  Thick lines -- recommended 1~km~s$^{-1}$ weak target
    catastrophic disruption criteria.
    \label{fig:compqstar}}}
\end{figure}

\end{document}